\documentclass[twocolumn, twocolappendix]{aastex631}
\usepackage{float}
\usepackage{wasysym}
\usepackage{booktabs}
\usepackage{changepage}
\usepackage{xspace}
\usepackage[encapsulated]{CJK}
\usepackage{savesym}
\savesymbol{iint}
\savesymbol{iiint}
\usepackage{amsmath}	
\restoresymbol{ams}{iint}
\restoresymbol{ams}{iiint}
\usepackage{amssymb}
\usepackage{multirow}
\usepackage{longtable}

\begin{document}

\begin{abstract}

In its first two years of operation, the James Webb Space Telescope has enabled the discovery of a surprising number of UV-bright galaxies at $z\sim10-14$. Their number density is still relatively uncertain, due to cosmic variance effects, and the limited survey area with deep imaging. Here, we combine pure parallel imaging from the PANORAMIC survey with data from legacy fields to constrain the bright end (M$_{\rm UV}<-18.5$) of the UV luminosity function (UVLF) over $0.28\,$deg$^2$ of NIRCam imaging in 6 or more filters, and along 35 independent lines of sight. Using conservative color selections, we compile robust dropout samples at $z\sim10$, $z\sim13$, and $z\sim17$, and identify 16 new candidates from PANORAMIC. Our inferred UVLFs at $z\sim10$ are consistent with literature results and we confirm the high abundance of galaxies at the bright end (M$_{\rm UV}\lesssim-21$) with better number statistics. We find somewhat lower number densities at $z\sim13$ compared to previous studies, and no robust candidates at $z\sim17$, indicating a rapid evolution of the galaxy population from $z\sim10-17$. The improved upper limits at $z\sim17$ imply that the cosmic UV luminosity density drops by at least a factor $\sim50$ from $z\sim10$ to $z\sim17$. Comparing our results to models proposed to explain the abundance of UV-bright galaxies at $z\gtrsim10$, we conclude that a modest increase in the star formation efficiency, or in the burstiness of star formation, a more top-heavy initial mass function, a lack of dust attenuation, or a combination of these effects at $z\gtrsim10$, is sufficient to match our observational constraints.

\end{abstract}


\title[PANORAMIC UVLF]{Exploring Cosmic Dawn with PANORAMIC I: The Bright End of the UVLF at $\mathbf{z\sim9 -17}$}

\author[0000-0001-8928-4465]{Andrea Weibel}
\affiliation{Department of Astronomy, University of Geneva, Chemin Pegasi 51, 1290 Versoix, Switzerland}

\author[0000-0001-5851-6649]{Pascal A.\ Oesch}
\affiliation{Department of Astronomy, University of Geneva, Chemin Pegasi 51, 1290 Versoix, Switzerland}
\affiliation{Cosmic Dawn Center (DAWN), Denmark}
\affiliation{Niels Bohr Institute, University of Copenhagen, Jagtvej 128, K{\o}benhavn N, DK-2200, Denmark}

\author[0000-0003-2919-7495]{Christina C.\ Williams}
\affiliation{NSF–DOE Vera C. Rubin Observatory/NSF NOIRLab, 950 N. Cherry Ave., Tucson, AZ 85719, USA}
\affiliation{Steward Observatory, University of Arizona, 933 North Cherry Avenue, Tucson, AZ 85721, USA}

\author[0000-0002-8896-6496]{Christian Kragh Jespersen}
\affiliation{Department of Astrophysical Sciences, Princeton University, Princeton, NJ 08544, USA}

\author[0000-0002-7087-0701]{Marko Shuntov}
\affiliation{Cosmic Dawn Center (DAWN), Denmark}
\affiliation{Niels Bohr Institute, University of Copenhagen, Jagtvej 128, K{\o}benhavn N, DK-2200, Denmark}
\affiliation{Department of Astronomy, University of Geneva, Chemin Pegasi 51, 1290 Versoix, Switzerland}

\author[0000-0001-7160-3632]{Katherine E. Whitaker}
\affiliation{Department of Astronomy, University of Massachusetts, Amherst, MA 01003, USA}
\affiliation{Cosmic Dawn Center (DAWN), Denmark}

\author[0000-0002-7570-0824]{Hakim Atek}
\affiliation{Institut d’Astrophysique de Paris, CNRS, Sorbonne Université, 98bis Boulevard Arago, 75014 Paris, France}

\author[0000-0001-5063-8254]{Rachel Bezanson}
\affiliation{Department of Physics and Astronomy and PITT PACC, University of Pittsburgh, Pittsburgh, PA 15260, USA}

\author[0000-0003-2680-005X]{Gabriel Brammer}
\affiliation{Cosmic Dawn Center (DAWN), Denmark}
\affiliation{Niels Bohr Institute, University of Copenhagen, Jagtvej 128, K{\o}benhavn N, DK-2200, Denmark}

\author[0009-0009-9795-6167]{Iryna Chemerynska} 
\affiliation{Institut d'Astrophysique de Paris, CNRS, Sorbonne Universit\'e, 98bis Boulevard Arago, 75014, Paris, France}

\author[0000-0001-9978-2601]{Aidan P. Cloonan}
\affiliation{Department of Astronomy, University of Massachusetts, Amherst, MA 01003, USA}

\author[0000-0001-8460-1564]{Pratika Dayal}
\affiliation{Kapteyn Astronomical Institute, University of Groningen, PO Box 800, 9700 AV Groningen, The Netherlands}
\affiliation{Canadian Institute for Theoretical Astrophysics, 60 St George St, University of Toronto, Toronto, ON M5S 3H8, Canada}

\author[0000-0001-6278-032X]{Lukas J. Furtak}
\affiliation{Physics Department, Ben-Gurion University of the Negev, P.O. Box 653, Be’er-Sheva 84105, Israel}

\author[0000-0003-3760-461X]{Anne Hutter}
\affiliation{Department of Astrophysics, University of Vienna, T\"urkenschanzstra{\ss}e 17, 1180 Vienna, Austria}

\author[0000-0001-7673-2257]{Zhiyuan Ji}
\affiliation{Steward Observatory, University of Arizona, 933 North Cherry Avenue, Tucson, AZ 85721, USA}

\author[0000-0003-0695-4414]{Michael V. Maseda} 
\affiliation{Department of Astronomy, University of Wisconsin-Madison, Madison, WI 53706, USA}

\author[0000-0003-1207-5344]{Mengyuan Xiao}
\affiliation{Department of Astronomy, University of Geneva, Chemin Pegasi 51, 1290 Versoix, Switzerland}

\section{Introduction}
\label{sec:intro}

Characterizing the build-up of the first galaxies that ended the Dark Ages and illuminated the Universe for the first time is one of the key goals of the James Webb Space Telescope \citep[JWST;][]{Gardner2023}. Prior to the era of JWST observations, the study of the highest redshift galaxies was limited by the wavelength coverage of the Wide Field Camera 3 (WFC3) onboard the Hubble Space Telescope (HST), probing $\lambda_{\rm obs}\lesssim1.6\,\mu m$. In principle, it enabled the detection of galaxies out to redshifts of $z\sim12$, at which point the Lyman break starts shifting out of even the reddest WFC3 filter, the H-band (F160W). \citet{Bouwens2011} identified a candidate galaxy at $z>10$, whose high redshift nature was debated in the literature \citep{Ellis2013,Brammer2013}, and that was confirmed to be at $z_{\rm spec}=11.58\pm0.05$ with JWST/NIRSpec in \citet{CurtisLake2023}. Pre-JWST candidates for even higher redshift galaxies based on ground-based and Spitzer/IRAC near-infrared imaging searches \citep{Harikane2022} turned out to be at $z<5$ with JWST spectroscopy \citep{Harikane2025}. The highest redshift galaxy with a Lyman break detected in HST slitless grism spectroscopy was GN-z11 \citep{Oesch2016}, later found to be at $z=10.6$ with JWST \citep{Bunker2023}. However, only a small number of robust galaxy candidates were known at $z\gtrsim9$, making determinations of the UV luminosity function (UVLF) challenging, with some papers suggesting a rapid decline in the number density of objects beyond $z\sim9$ \citep[e.g.][]{Bouwens2016, Ishigaki2018, Oesch2018}, and others finding a more shallow evolution \citep[e.g.][]{McLeod2016}. With the NIRCam instrument mounted on JWST \citep{Rieke2023}, the accessible wavelength range has been extended to $\sim 5\,\mu m$ at comparable spatial resolution and sensitivity, meaning that Lyman Break Galaxies (LBGs) of similar apparent magnitudes can in principle be identified out to $z>20$. 

Not only has the redshift record been broken multiple times in the first three years of JWST observations, with spectroscopic confirmations now out to $z\sim14$ \citep{CurtisLake2023, Carniani2024,Naidu2025b}, but JWST has also revealed a surprisingly high number of luminous $z\gtrsim10$ galaxies compared to extrapolations of lower redshift UVLFs, and theoretical expectations \citep[e.g.][]{Dayal2014, Mason2015, Tacchella2018, Williams2018, Behroozi2019}. These galaxies were first identified as candidates from imaging data \citep[e.g.][]{Naidu2022,Castellano2022, Adams2023,Finkelstein2023,Atek2023,Casey2024,Hainline2024}, and some were later confirmed spectroscopically \citep[e.g.][]{ArrabalHaro2023b,Castellano2024,Fujimoto2024,Harikane2024}. 

As a consequence, recently derived UVLFs at $z\sim10-15$ show an excess at the bright end, and indicate a shallow evolution of the UV luminosity density with respect to pre-JWST expectations \citep[e.g.][]{Donnan2023,PerezGonzalez2023b,Harikane2023,Adams2024,Finkelstein2024,Robertson2024,Donnan2024,Whitler2025}. While some photometric candidates contributing to these UVLFs still await spectroscopic confirmation, the high number density at the bright end of the UVLF is seen from sources with spectroscopic redshifts only out to $z\sim13$ \citep{Harikane2024}. This finding has sparked an extensive debate on its possible causes in the literature. Frequently discussed theoretical explanations include higher star formation efficiencies at high redshift \citep{Dekel2023,Li2024,BoylanKolchin2024,Mauerhofer2025}, a lack of dust attenuation \citep{Ferrara2023}, bursty star formation histories \citep{Mason2023,Shen2023,Sun2023,Kravtsov2024}, and a more top heavy initial mass function (IMF) with increasing redshift and/or in highly star-forming regions \citep{Yung2024, Trinca2024, Cueto2024, Lu2025, Mauerhofer2025, Hutter2025}. Other authors also explored helium enhancement and nebular continuum emission \citep{Katz2024b}, a contribution of AGN to the observed UV-luminosity \citep{Pacucci2022,Hegde2024}, or larger dust grain sizes \citep{Toyouchi2025,Narayanan2025} as possible drivers of the observed abundance of UV-bright galaxies at $z\gtrsim10$.

Despite all this ongoing discussion, galaxy sample sizes remain small at the bright end of the UVLF at $z\gtrsim10$ due to the limited survey area that has been imaged with JWST at sufficient depth and wavelength resolution. Further, most determinations of the UVLF based on early JWST data rely on a small number of independent lines of sight, meaning that they are subject to significant cosmic variance, i.e., variations in the measured number density of galaxies due to the clustering of dark matter halos and galaxies in the cosmic web \citep[e.g.][]{Moster2011,Jespersen2025}. Indeed, studies based on imaging along a larger number of independent sightlines somewhat mitigated earlier findings. \citet{Willott2024} analysed data along 5 independent lines of sight from the CAnadian NIRISS Unbiased Cluster Survey (CANUCS, \citealt{Willott2022}), and found a much steeper decline in the space density of UV-selected galaxies from $z\sim8-12$ relative to other early JWST studies, with no candidates at $z>12.5$. Using the first 19 fields from the pure-parallel program Bias-free Extragalactic Analysis for Cosmic Origins with NIRCam (BEACON), \citet{Morishita2025} found moderate number densities of galaxies at $z\sim10-13$ compared to previous results, and no robust candidates at $z>13$ resulting in upper limits that are just consistent or even below previous determinations of the UVLF. Cosmic variance effects are also emphasized in a recent paper from \citet{Asada2025} who derived the UVLF at $10<z<16$ from deep medium and wide-band imaging in three independent NIRCam-pointings from CANUCS and the JWST in Technicolor program (GO-3362, PI Muzzin) and found that it is 0.6 dex lower than the UVLF measured in the JADES Origins Field with similar NIRCam coverage \citep{Robertson2024}. Complementarily, \citet{Rojas-Ruiz2024} used a sample of spectroscopically confirmed galaxies from the BoRG-JWST survey \citep{Roberts-Borsani2024b}, previously identified from pure-parallel HST-imaging along $>200$ independent lines of sight, and found higher number densities of UV-luminous galaxies at $z\sim7-9$, compared to estimates based on JWST legacy fields, which may alleviate the implied shallow evolution of the bright end of the UVLF with redshift out to $z\gtrsim10$. 

These results illustrate the need for a large number of independent lines of sight to obtain an unbiased view of the evolution of the UVLF and the build-up of galaxies at Cosmic Dawn \citep[see e.g.,][]{Trapp2022}. Such data can be provided by JWST/NIRCam pure parallel imaging, as collected in Cycle 1 by the Parallel wide-Area Nircam Observations to Reveal And Measure the Invisible Cosmos (PANORAMIC, \citealt{Williams2025}).

Efficient selection of high redshift galaxies through the Lyman break technique \citep[e.g.][]{Steidel1996, Madau1996} is constrained to redshift ranges that are feasible for the photometric filters of a given survey. In the case of PANORAMIC imaging, those are the short-wavelength (SW) channel filters F115W, F150W, and F200W, probing wavelengths of $\sim1-2\mu m$. While a large number of high-redshift candidates have been identified and confirmed through F115W and F150W dropouts ($z\sim9-11$ and $z\sim12-14$ respectively), F200W dropouts remain rare, and some seemingly robust candidates have turned out to be transients or lower redshift interlopers. For example, a $z\sim16$ candidate from \citet[][with weak detected flux in F200W]{Hainline2024}, has turned out to be a supernova \citep{DeCoursey2025}. The most prominent example, however, is the so-called Schr\"odinger galaxy. First identified as a bright and robust $z\sim16.6$ candidate in CEERS imaging \citep{Naidu2022, Harikane2023, Donnan2023}, this source turned out to be a $z=4.9$ dusty starburst galaxy \citep{ArrabalHaro2023b}, as already suggested in \citet{Naidu2022b} and \citet{Zavala2023}.

Recently, candidate galaxies at $z\gtrsim15$ have been identified in the deepest available JWST imaging over small fields \citep{Kokorev2025,PerezGonzalez2025}. These candidates are relatively faint (M$_{\rm UV}\gtrsim-19$), making their spectroscopic confirmation challenging. \citet{Castellano2025} performed a careful search for $z\gtrsim15$ galaxies over $\sim0.2{\rm deg}^2$ of legacy imaging, and identified 9 brighter candidates (M$_{\rm UV}\lesssim-19$) at $z\sim15-20$. However, as they discuss, two additional sources that satisfy their selection criteria, and have NIRSpec spectroscopy turn out to be low-redshift interlopers, and different photometric redshift codes identify secondary low-redshift solutions for all of their candidates.

Here, we use imaging data from PANORAMIC, in combination with publicly available NIRCam imaging in various legacy fields to select F115W, F150W, and F200W dropouts and estimate the bright end of the UVLF at $z\gtrsim9$ based on 35 independent lines of sight covering a total area of $1020\,$arcmin$^2$, or $0.28\,$deg$^2$ -- the largest survey area with deep JWST imaging data in 6 or more filters used to construct a UVLF to date. 
We further leverage the large number of independent sightlines to perform a measurement of cosmic variance and through it the galaxy bias at $z\sim10$ in a companion paper \citep{Weibel2025}.

This paper is structured as follows. We present an overview of the imaging data, describe the photometric catalogs, the sample selection, the spectral energy distribution (SED) fitting, as well as our determination of sample completeness and our calculation of the UVLF with its uncertainties in Section \ref{sec:methods}. In Section \ref{sec:results}, we present our results, starting with a discussion of each of the three samples and some new and outstanding candidates, followed by a comparison of our UVLFs and the inferred UV luminosity density with the literature, as well as various models and simulations. We discuss our findings in Section \ref{sec:discussion}, and conclude in Section \ref{sec:summary_conclusions}.

Throughout this work, we assume a $\Lambda$CDM cosmology with parameters from the nine-year Wilkinson Microwave Anisotropy Probe Observations \citep{WMAP9}, $h=0.6932$ and $\Omega_{m,0}$=0.2865. All magnitudes are reported in the AB system. 

\section{Data and Methods}
\label{sec:methods}

\subsection{Imaging}
\label{sec:imaging}

The basis of this work are the publicly available imaging data from JWST and HST from various legacy surveys, as well as the PANORAMIC pure parallel data. All imaging mosaics have been consistently reduced through the \texttt{grizli} software package \citep{grizli}, starting from level 2 calibrated data products from the Mikulski Archive for Space Telescopes (MAST), as outlined in e.g., \citet{Valentino2023}. We use JWST imaging data in the following fields: (1) in the EGS with data from the Cosmic Evolution Early Release Science Survey (CEERS, ERS-1345, PI Finkelstein, \citealt{Finkelstein2025}), GO-2279 (PI Naidu), DDT-2750 (PI Arrabal-Haro), and GO-2234 (PI Banados); (2) in the UDS and COSMOS fields with data from the Public Release Imaging for Extragalactic Research (PRIMER, GO-1837, PI Dunlop, see, e.g., \citealt{Donnan2024}), COSMOS-Web (GO-1727, PIs Kartaltepe \& Casey, \citealt{Casey2023}), GO-1810 (PI Belli), GO-1840 (PI Alvarez-Marquez), and DDT-6585 (PI Coulter); (3) in the GOODS fields (North and South) with data from the JWST Advanced Deep Extragalactic Survey (JADES, GTO-1180 \& GTO-1181, PI Eisenstein, as well as GTO-1210, PI Luetzgendorf, \citealt{Eisenstein2023a}), First Reionization Epoch Spectroscopically Complete Observations (FRESCO, GO-1895, PI Oesch, \citealt{Oesch2023}), Complete NIRCam Grism Redshift Survey (CONGRESS, GO-3577, PIs Egami \& Sun), GO-2079 (PI Finkelstein), the JWST Extragalactic Medium-band Survey (JEMS, GO-1963, PIs Williams, Maseda \& Tacchella, \citealt{Williams2023}), the JADES Origins Field (GO-3215, PI Eisenstein, \citealt{Eisenstein2023b}), GTO-1264 (PI Robledo), GO-4762 (PI Fujimoto), and Slitless Areal Pure-Parallel High-Redshift Emission Survey (SAPPHIRES, GO-6434, PI Egami); (4) the Abell-2744 cluster with data from Ultra-deep NIRCam and NIRSpec Observations Before the Epoch of Reionization (UNCOVER, PIs Labb\'e \& Bezanson, \citealt{Bezanson2024}), GLASS (ERS-1324, PI Treu, \citealt{Treu2022}), DDT-2756 (PI Chen),  All the Little Things (ALT, GO-3516, PIs Naidu \& Matthee, \citealt{Naidu2024}), and Medium bands, Mega Science (GO-4111, PI Suess, \citealt{Suess2024}); (5) the JWST North Ecliptic Time-Domain Field with data from Prime Extragalactic Areas for Reionization and Lensing Science (PEARLS, GTO-2738, PIs Windhorst \& Hammel, \citealt{Windhorst2023}).

As an important extension, we include data from the PANORAMIC survey (GO-2514, PIs Williams \& Oesch, \citealt{Williams2025}), grouped in 39 associations (i.e., imaging mosaics). Most of these contain a single NIRCam pointing with imaging in six wide filters (F115W, F150W, F200W, F277W, F356W, and F444W) with a few deep pointings having additional data in the F410M band, and a few associations containing multiple pointings. Of the 39 associations, 9 overlap with or are in close proximity to the UDS, COSMOS or one of the GOODS fields, adding depth and/or area to those legacy fields. For full details on the survey design and data reduction of PANORAMIC, see \citet{Williams2025}. In total, PANORAMIC provides imaging in 6 or more filters over $432\,$arcmin$^{2}$, of which $\sim250\,$arcmin$^2$ are split over 28 independent and new lines of sight (at $>2\,$deg separation from other fields).

In addition, the cycle 2 pure parallel program BEACON (GO-3990, PIs Morishita, Mason, Treu \& Trenti, \citealt{Morishita2025}) contributes imaging to the mosaics in the EGS, UDS, and Abell-2744 fields.

We use version 7.0 or higher imaging products from the DAWN JWST Archive (DJA)\footnote{\url{https://dawn-cph.github.io/dja/imaging/v7/}} in all the legacy fields, as well as version 7.2 mosaics from the first PANORAMIC data release\footnote{\url{https://panoramic-jwst.github.io/}} which can be found as high-level science products on MAST\footnote{\dataset[doi:10.17909/fpzr-as35
]{https://doi.org/10.17909/fpzr-as35}}.

We additionally modify the science mosaics in the Abell-2744 field to remove the brightest cluster galaxies (BCGs). To this end, we compute a running median filter with a box-size of $101\times101$ pixels ($4.04$\arcsec $\times$ 4.04\arcsec) and subtract it from the original images.

All the NIRCam long-wavelength (LW) channel and HST filters are drizzled to a pixel scale of 0.04\arcsec\ while the NIRCam SW channel filters are usually drizzled to a smaller pixel scale of 0.02\arcsec\ to leverage their higher spatial resolution. Exceptions are the large CEERS and PRIMER mosaics where all filters are drizzled to 0.04\arcsec.

\subsection{Photometric Catalogs}
\label{sec:catalogs}

From the uniformly reduced imaging mosaics we generate point spread function (PSF) matched photometric catalogs, following the methods outlined in \citet{Weibel2024} and \citet{Williams2025}. To summarize briefly, we generate inverse-variance weighted stacks of the F277W, F356W, and F444W mosaics in each field, and run \texttt{SourceExtractor} \citep{Bertin1996} in dual mode, using the stacks as the detection image, and measuring fluxes through circular apertures with 0.16\arcsec\ radius on PSF matched versions of all  available JWST and HST mosaics. These aperture fluxes are scaled to total based on a Kron ellipse determined on the detection image and an additional correction to account for flux in the wings of the PSF, beyond the Kron ellipse.

For most filters in the legacy fields, PSFs are extracted from the images using an automatically generated list of stars and the python tool  \texttt{psf.EPSFBuilder} \citep{Anderson2000, Anderson2016} which is part of the \texttt{photutils} package \citep{Bradley2022}. However, some filters only cover a small area within a given legacy field, e.g., a single NIRCam pointing ($\sim9.7\,$arcmin$^2$) which usually does not contain a sufficient number of isolated stars for the construction of an empirical PSF. In such cases, as well as for all the PANORAMIC pointings (most of which are single NIRCam pointings), we instead use \texttt{webbpsf} \citep{Perrin2014} with a \texttt{jitter\_sigma} parameter of 0.02 to derive PSFs, which we further rotate to match the position angle of each observation.

Whenever NIRCam/SW mosaics are available on a 0.02\arcsec\ pixel scale, we re-bin those mosaics to 0.04\arcsec\ since \texttt{SourceExtractor} requires all mosaics to be on the same pixel grid. However, we extract PSFs from the 0.02\arcsec\ images and then re-bin the PSFs to benefit from the more accurate sampling of the strongly centrally peaked radial profiles.

\subsection{Sample Selection}
\label{sec:sample_selection}

We proceed with the sample selection, which is mainly based on photometric fluxes and colors, rather than photometric redshifts, simplifying the subsequent treatment of completeness (Section \ref{sec:completeness}). Given that we wish to probe the bright end of the UVLF, we choose conservative selection cuts with the goal of obtaining a pure sample that avoids significant contamination of low redshift interlopers.

Our three redshift bins are determined by the wavelength coverage of the NIRCam/SW filters at $\lambda>1\,\mu m$: F115W, F150W, and F200W. We select dropouts in each of these filters as follows:

\begin{gather}
    {\rm \mathbf{F115W\,\,dropouts}}\,\,\mathbf{(z\sim10)}\nonumber\\[1ex]
    {\rm F115W} - {\rm F150W} > 1.5\,\,\,\,\land\nonumber\\[1ex]
    {\rm F150W} - {\rm F356W} < 0.5\,\,\,\,\land\\[1ex]
    {\rm SNR(F150W)} > 8\,\,\,\,\land\nonumber\\[1ex]
    {\rm SNR(F200W)} > 3\nonumber
\end{gather}

\begin{gather}
    {\rm \mathbf{F150W\,\,dropouts}}\,\,\mathbf{(z\sim13)}\nonumber\\[1ex]
    {\rm F150W} - {\rm F200W} > 1.5\,\,\,\,\land\nonumber\\[1ex]
    {\rm F200W} - {\rm F444W} < 0.5\,\,\,\,\land\\[1ex]
    {\rm SNR(F200W)} > 8\,\,\,\,\land\nonumber\\[1ex]
    {\rm SNR(F277W)} > 3\nonumber
\end{gather}

\begin{gather}
    {\rm \mathbf{F200W\,\,dropouts}}\,\,\mathbf{(z\sim17)}\nonumber\\[1ex]
    {\rm F200W} - {\rm F277W} > 1.5\,\,\,\,\land\nonumber\\[1ex]
    {\rm F277W} - {\rm F444W} < 0.5\,\,\,\,\land\\[1ex]
    {\rm SNR(F277W)} > 8\,\,\,\,\land\nonumber\\[1ex]
    {\rm SNR(F356W)} > 3\nonumber
\end{gather}

where FXXXW refers to the AB-magnitude in the respective filter, and we replace fluxes with $<2\sigma$ detections by $2\sigma$ upper limits before converting to AB-magnitudes. The second SNR-cut (e.g., SNR(F200W$)>3$ in the F115W dropout selection) is to remove sources that only have a high SNR in one filter (typically diffraction spikes or other spurious detections) without biasing against true high-z galaxies with blue SEDs. We further apply the \texttt{use\_phot} flag from \citet{Weibel2024} which removes a small number of spurious objects and sources that were flagged as stars.

Three reviewers then visually inspect two sets of NIRCam imaging cutouts with stronger and weaker contrast as well as the photometric fluxes of all candidates. We do not inspect any SED-fitting output to avoid being influenced by the preferred solution of an SED-fitting code, and we mainly remove artifacts (e.g., diffraction spikes and hot pixels), and sources with image quality issues or contaminated photometry. We further remove objects with clearly visible flux in filters probing wavelengths below the supposed Lyman break, usually the NIRCam filter just blueward of the break, i.e., F150W, F115W, and F090W for the F200W, F150W, and F115W dropouts. We note here that F090W is not available in all fields, and refer to a more general discussion of the effect of the inhomogeneous filter coverage across fields in Section \ref{sec:inhomogeneous_coverage}.

During visual inspection, we noticed a number of candidates that nominally satisfy the selection cuts, and are real sources with unproblematic photometry, but whose SED is either too blue to be fit with a reasonable UV-continuum, or suspiciously red, in combination with a smooth turn-over around the supposed Lyman break. These sources can be reliably identified to be emission line or Balmer break galaxies at $z\lesssim5$ through SED-fitting. We therefore run the template-based SED-fitting code \texttt{eazy} \citep{Brammer2008} to estimate photometric redshifts using the \texttt{blue\_sfhz} template set\footnote{\url{https://github.com/gbrammer/eazy-photoz/tree/master/templates/sfhz}} which consists of 13 templates with redshift-dependent properties generated through the Flexible Stellar Population Synthesis (\texttt{FSPS}) code \citep{Conroy2009, Conroy2010} and a 14th template based on a strong emission line galaxy at $z=8.5$ from \citet{Carnall2023a}. We let the redshift vary freely in the range $z\in(0.01,20)$. In order not to significantly bias our color selection, we only remove sources that are fit at low redshift with very high confidence. Specifically, we remove sources with P$(z<8.5)>0.95$, i.e., with a $>$95\% probability of being at a redshift $<8.5$ according to the \texttt{eazy} redshift probability distribution.

In the F115W dropout sample, we selected two out of three images of the triply imaged $z\sim10$ candidate first published in \citet{Zitrin2014}, and with spectroscopic redshifts from UNCOVER and the DDT program DD-2756 (PI Chen, \citealt{Chen2022}). We remove the fainter of the two images to avoid double-counting the source and are left with final sample sizes of 86 F115W dropouts, 11 F150W dropouts and 4 F200W dropouts.

Further, 10 objects in total (6 F115W dropouts, 3 F150W dropouts and 1 F200W dropout) have aperture correction factors $>3$ (see Section \ref{sec:catalogs}). In all cases, this is due to the Kron ellipse being contaminated by a bright neighboring galaxy. We re-measure the total fluxes of those 10 objects by multiplying their aperture fluxes by the median aperture correction and dividing by the median encircled energy of the Kron ellipse on the F444W PSF measured from all our sample galaxies, resulting in a total correction factor of 1.87. While the resulting total fluxes may not be accurate depending on the morphology of the sources, we argue this effect should be minimal as both the affected galaxies and most of the sample are compact.

Finally, we determine the nominal redshift range of each sample based on the NIRCam filter throughput curves. We assume a flat SED in f$_\nu$ ($\beta=-2$), with a sharp Lyman break, i.e., f$_\nu(\lambda)={\rm C}=const.$, if $\lambda_{\rm rest}>1216$\AA\, and f$_\nu=0$ if $\lambda_{\rm rest}<1216$\AA. Shifting this SED to higher and higher redshift, we measure the synthetic fluxes in the NIRCam bands and determine the redshift at which the flux in the first dropout filter, F115W, drops to ${\rm C}/2$, due to the Lyman break entering that filter. This marks the lower end of our lowest redshift bin. Nominally, our selection with a color ${\rm F115W}-{\rm F150W}>1.5$ corresponds to a factor of $\sim4$ in $f_\nu$, but the factor of 2 generously accounts for noise in both filters. Shifting the synthetic SED to even higher redshifts, the Lyman break starts to enter F150W, decreasing the measured color. We then determine the redshift at which the flux in F150W drops to ${\rm C}/2$ as both the upper bound on the lowest redshift bin, and the lower bound of the second redshift bin (F150W dropouts), and repeat this process for the F200W dropouts. This yields nominal ranges of $8.6<z<11.3$, $11.3<z<15.3$, and $15.3<z<21.9$ for the F115W, F150W, and F200W dropout samples respectively.

\subsection{SED-Fitting}
\label{sec:sed_fitting}

To estimate the photometric redshifts, UV-magnitudes M$_{\rm UV}$ and -slopes $\beta$ of our sample galaxies, we run the Bayesian Analysis of Galaxies for Physical Inference and Parameter EStimation tool (\texttt{bagpipes}; \citealt{Carnall2018}), constraining the redshift to be in the respective nominal range determined above, adding (subtracting) 0.1 at the upper (lower) end of the bin to allow for some additional flexibility.

We further assume the non-parametric star formation history (SFH) model introduced in \citet{Leja2019} with the bursty continuity prior as described in \citet{Tacchella2022} to account for supposed bursty star formation at high redshifts \citep[e.g.,][]{Ciesla2024, Cole2023, Looser2023}. The SFH is split into 8 time-bins of which the first three are fixed to $0-10$, $10-30$, and $30-70\,$Myr lookback time. The other five bins are set to be equally sized in linear time from $70\,$Myr lookback time to $z=30$.
We use a uniform prior on the metallicity in the range $Z_*\in(0.01, 1)\,{\rm Z_\odot}$.

Our implementation of dust is similar to the setup described in \citet{Carnall2023c}. It is based on the model of \citet{Salim2018} with a uniform prior on the absolute attenuation in the V-band, $A_V\in(0,1)$. We further fit for the deviation from the \citet{Calzetti2000} slope, $\delta\in(-0.3,0.3)$ with a Gaussian prior centered at 0 and with a dispersion of 0.1, for the strength of the UV-bump at 2175\,\AA\, $B\in(0,5)$ with a uniform prior, and for the multiplicative factor on $A_V$ for stars in birth clouds (defined to be stars younger than $20\,$Myr.), $\eta\in(1,3)$, with a uniform prior. While these dust properties are likely degenerate with the SFH and hard to constrain from rest-frame UV photometry, we prefer to allow for flexibility in the fits, and marginalize over possible dust effects on the UV-continuum to get conservative estimates on the uncertainty in M$_{\rm UV}$ and $\beta$ which, apart from the redshift, are the only SED-fitting derived quantities used in this work.

The stellar population models used in the fitting are from \texttt{BPASS-v2.2.1} \citep{BPASS} which assumes a broken power-law IMF as described in \citet{Eldridge2017}, building on \citet{Kroupa1993}. Nebular emission in \texttt{bagpipes} is modeled with \texttt{CLOUDY} \citep{Ferland2017} as described in \citet{Carnall2018}, and we fit for the logarithm of the ionization parameter in the range log(U)$\in(-4,-1)$ assuming a uniform prior.

\texttt{bagpipes} then allows us to measure the UV-properties of each sample galaxy in a Bayesian way. To this end, we first sample 500 realizations of the fitted SED from the posterior distribution. For each realization, we keep track of the fitted redshift, perform a linear regression fit in the range 1350\AA\ $<\lambda_{\rm rest}<$ 2800\AA\ to infer the UV-slope $\beta$, and measure the UV-magnitude M$_{\rm UV}$ as the mean flux in the range 1450\AA\ $<\lambda_{\rm rest}<$ 1550\AA. The median and percentiles of the 500 measurements respectively then constitute our final values and uncertainties of M$_{\rm UV}$, $\beta$, and $z$.

\subsection{Spectroscopic Redshifts}
\label{sec:z_spec}

\begin{figure}
     \centering
     \includegraphics[width=0.47\textwidth]{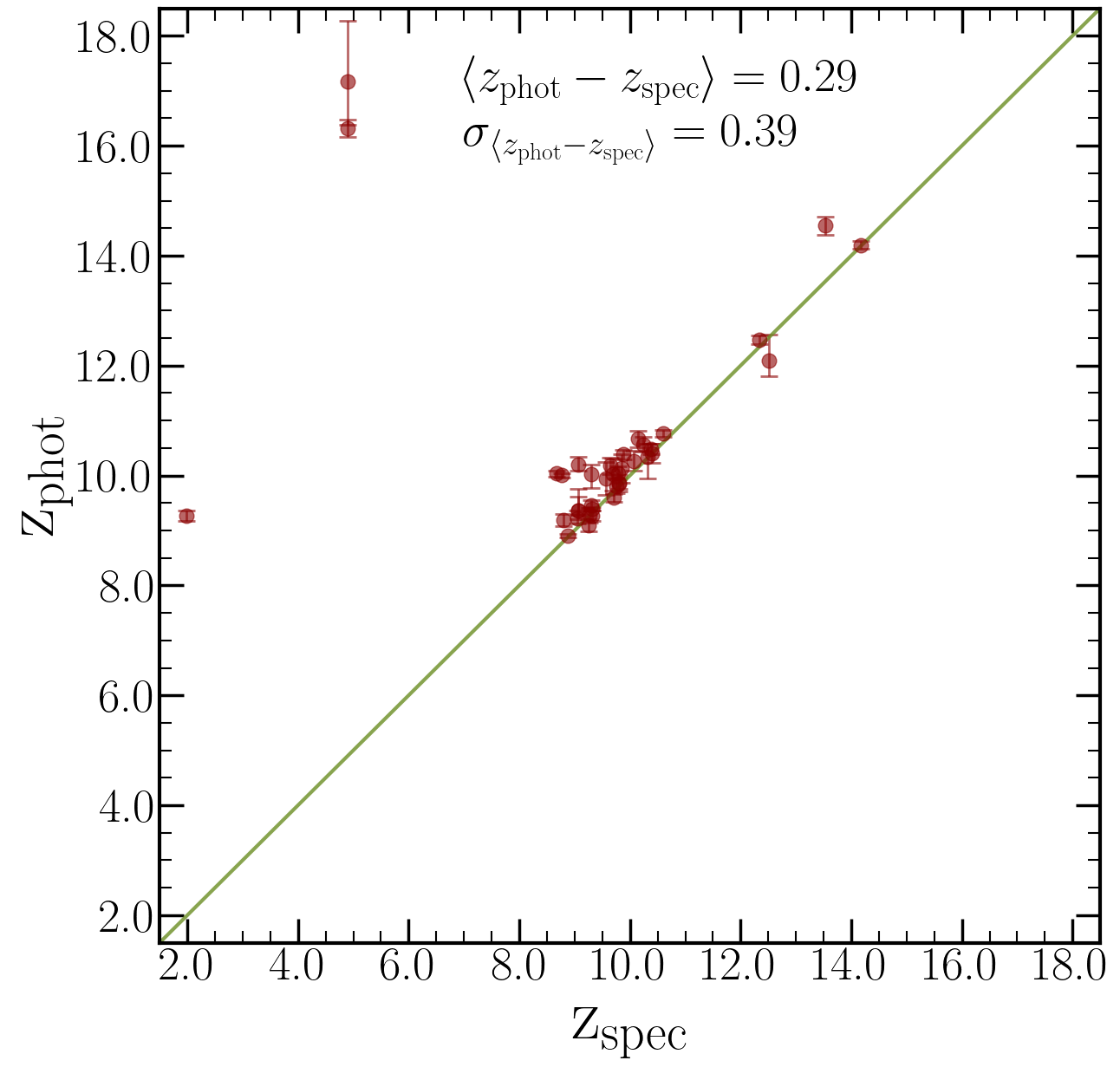}
     \caption{Comparison of photometric with spectroscopic redshifts for the 37 galaxies that have a grade 3 (i.e., high-quality) spectrum on the DJA, as well as PAN-z14-1 with its spectroscopic redshift from \citet{Donnan2026}. We find that the $z_{\rm phot}$ estimates from \texttt{bagpipes} are in good agreement with $z_{\rm spec}$, with only three strong outliers, two in the F200W dropout sample that are $z=4.9$ strong emission line sources (see Section \ref{sec:z17_sample}), and one in the F115W dropout sample which may be a high redshift galaxy that is blended with a low redshift interloper \citep{Bunker2024}.}
     \label{fig:zphot_vs_zspec}
 \end{figure}

We match our dropout samples with all grade 3 spectra (robust redshifts confirmed through visual inspection) on the DJA available as of January 22, 2026, finding 32, 3, and 2 matches in the three samples. In addition, the source with ID 35840 in the PANORAMIC pointing J221700P0025 has been confirmed to lie at $z_{\rm spec}=13.53^{+0.05}_{-0.06}$ in \citet{Donnan2026}, and we add their measurement to the comparison (see also Section \ref{sec:z13_sample} and Figure \ref{fig:z13_candidates}). We plot the available spectroscopic redshifts against our photometric redshifts from \texttt{bagpipes} in Figure \ref{fig:zphot_vs_zspec}.

Remarkably, all except for three of our photometric candidates with available spectra are confirmed to be at high redshift. Two of the exceptions are F200W dropouts. One is the Schr\"odinger galaxy at $z_{\rm spec}=4.9$ \citep{ArrabalHaro2023b}, the other has a spectrum from the CANDELS-Area Prism Epoch of Reionization Survey (CAPERS, GO-6368, PI Dickinson) that also puts it at $z_{\rm spec}=4.9$. We discuss these sources further in Section \ref{sec:z17_sample}. The third outlier is an F115W dropout that has a spectrum from JADES (GTO-1210, PI Luetzgendorf) which puts it at $z\sim1.98$, but the spectrum may overlap with that of a high redshift galaxy in the background \citep{Bunker2024}.

With JWST measuring spectroscopic redshifts for dozens of galaxies at $z\gtrsim10$, photometric redshift estimates have been found to be biased high at those redshifts. This may be related to an Eddington-type bias as discussed in \citet{Serjeant2023}, or incorrect modeling of the strong Lyman-$\alpha$ damping wing caused by high neutral hydrogen column densities \citep{Heintz2025}. Indeed, we also see this effect with our \texttt{bagpipes} setup, finding a mean offset $\langle z_{\rm phot} - z_{\rm spec}\rangle = 0.29$ with a scatter of 0.39. We note that if we instead compare to photometric redshifts from \texttt{eazy}, as derived in Section \ref{sec:sample_selection}, these numbers reduce to 0.12 and 0.30 (see also \citealt{Hainline2026} investigating this effect for different template sets in \texttt{eazy}). We however prefer to derive all quantities self-consistently within \texttt{bagpipes}, without putting strong priors on the redshift, so as to obtain conservative estimates on the uncertainty of the derived UVLFs. Further, the impact of this bias on the derived values of M$_{\rm UV}$ is marginal, and it does not cause galaxies to scatter from one redshift bin to another. Therefore, its impact on the derived UVLFs is minimal.

\begin{figure}
     \centering
     \includegraphics[width=0.47\textwidth]{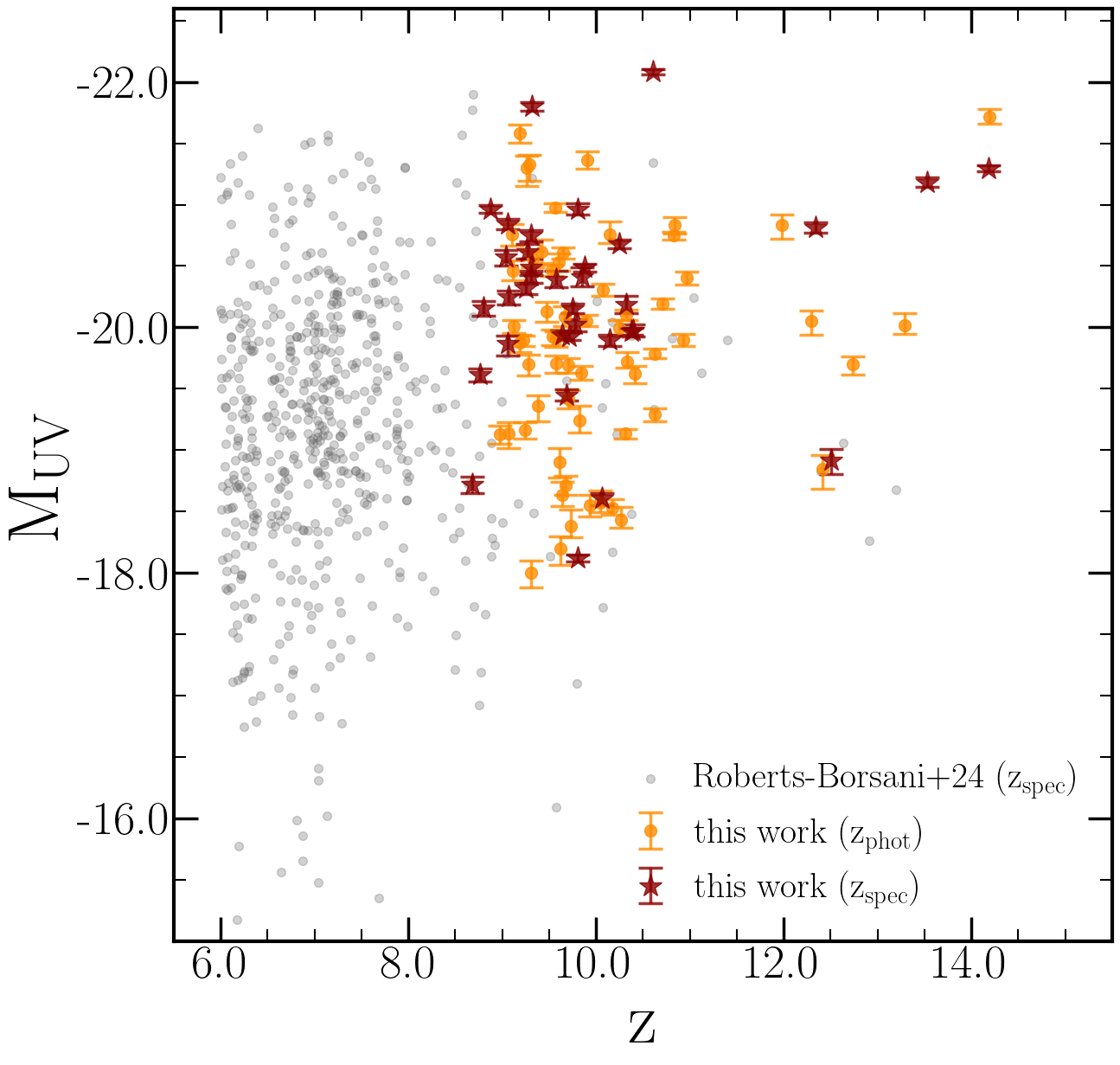}
     \caption{Absolute UV-magnitude M$_{\rm UV}$ against redshift for all the sample galaxies with (red stars) or without (orange circles) spectroscopic redshifts. This shows that sources with z$_{\rm spec}$ are not substantially biased towards the brightest objects, relative to our photometric sample. The grey markers are a compilation of spectroscopically confirmed galaxies from \citet{Roberts-Borsani2024} shown for reference.}
     \label{fig:muv_vs_z}
 \end{figure}

One might expect spectroscopically confirmed sources at $z\gtrsim9$ to be biased towards the brightest and therefore most robust photometric candidates that were selected from early JWST or HST imaging data, and prioritized in the NIRSpec/MSA mask design. Further, robust spectroscopic redshifts can only be measured from relatively bright sources and/or deep spectra. To test this, we plot M$_{\rm UV}$ against redshift for our entire sample of galaxies in Figure \ref{fig:muv_vs_z}, highlighting those with spectroscopic redshifts as red stars. A compilation of spectroscopically confirmed objects from \citet{Roberts-Borsani2024} is shown in the background. Note that this sample is only shown for reference and that it contains sources that are also part of our sample, but with slightly different measured redshifts and M$_{\rm UV}$ due to the independent data processing.

While the spectroscopically confirmed sample does contain some of the brightest sources in the sample at $z\sim10$ as well as the brightest source at $z\sim14$ \citep{Carniani2024,Carniani2024b}, the median M$_{\rm UV}$ of the sources with $z_{\rm spec}$ ($-20.3$) is only slightly lower than that of the sources without $z_{\rm spec}$ ($-19.9$), indicating that our spectroscopic sample is not substantially biased towards brighter sources. This may be due to the ``wedding-cake'' nature of the compilation of all the legacy surveys which enabled a robust selection and spectroscopic follow-up of bright high-z candidates from large areas, as well as fainter candidates from deeper imaging over smaller areas.

However, the spectroscopically confirmed sources have somewhat more robust photometric redshifts from \texttt{bagpipes} with a median 1$\sigma$ uncertainty of 0.13, compared to sources without z$_{\rm spec}$ that show a median 1$\sigma$ uncertainty of 0.23. This is in part because the spectroscopically confirmed sources mostly lie in legacy fields with relatively deep imaging data in many filters. We discuss this in more detail in Section \ref{sec:inhomogeneous_coverage}. The success rates of 97\%\ and 100\%\ of our F115W and F150W dropout selections among the spectroscopically confirmed objects therefore likely overestimates the purity of the sample. That said, it is nevertheless re-assuring and confirms that our rather conservative selection cuts successfully identify true high redshift galaxies with a low expected interloper fraction.

As an additional test of our selection, we cross-match our full photometric catalog with all grade 3 sources on the DJA with $z_{\rm spec}>8.6$, and find 91 confirmed high redshift galaxies that are not included in our sample. 78 of them would fall in our F115W dropout sample, but do not satisfy our Lyman break color selection and/or the SNR-cut in F150W. 36 of those sources lie at z$_{\rm spec}<9$, the lower end of the nominal redshift range of the F115W dropout selection. These sources often have measured Lyman break colors $<1.5$, even if they are relatively bright, because of the Lyman break being in the F115W filter. The other 42 sources are either at the upper end of the redshift range (16 sources at $z_{\rm spec}>10$), where the Lyman break enters F150W, or are too faint to pass our selection. We further find 11 sources that would fall in our F150W dropout sample. All of them have Lyman break colors $<1.5$ (10 cases) and/or SNR(F200W$)<8$ (6 cases). This means that the depth of the F150W and F200W imaging is insufficient to measure a color $>1.5$ and/or a ${\rm SNR}>8$. These effects are taken into account in our completeness correction (Section \ref{sec:selection_completeness}). Of the remaining 2 sources, one lies at $z_{\rm spec}=9.94$, and is not included here due to a red color of ${\rm F150W} - {\rm F356W}=0.6$, slightly above our threshold value of 0.5. The final source at $z_{\rm spec}=9.30$ was removed in our visual inspection due to flux in F090W which is likely spurious. This exercise illustrates that our selection is limited to bright sources and/or sufficiently deep photometry, and it highlights the power of NIRSpec to measure redshifts of relatively faint sources even at $z>10$.

\subsection{Gravitational Magnification in A2744}
\label{sec:lensing}

For the sources in the Abell-2744 field, we correct their measured UV-magnitudes for gravitational lensing magnification. We use the \texttt{v2.0} UNCOVER strong lensing model of the cluster, which was constructed by \citet{Furtak2023} using the \citet{Zitrin2015} analytic method and has recently been updated with new spectroscopic redshifts from JWST in \citet{Price2025}. This parametric model, comprising five smooth dark matter halos and 552 cluster galaxies, is constrained with 187 multiple images belonging to 66 sources, 60 of which have spectroscopic redshifts. The model achieves a final average image reproduction error of $\Delta_{\mathrm{RMS}}=0.6\arcsec$.

We compute magnifications at each source's position and redshift. Of the 12 (2) F115W (F150W) dropouts identified in the Abell-2744 field, 10 (2) have a low magnification in the range $\mu=1.1-2.2$. Among the F115W dropouts, there is one source with $\mu=4.1$ (ID 37210), and one with $\mu=12.2$ (ID 49489, the source from \citealt{Zitrin2014}). In order to calculate the volume effectively probed in the Abell-2744 field (see Section \ref{sec:survey_area}), we construct source planes for each redshift bin.

\subsection{Completeness}
\label{sec:completeness}

While we have shown that it is reasonable to assume that our samples have a high purity, we need to quantify their completeness in order to measure the UVLF. This is aided by the simplicity of our selection which mostly relies on photometric fluxes and colors. We assume here that the removal of confident low redshift interlopers based on \texttt{eazy} (see Section \ref{sec:sample_selection}) has no effect on the completeness. We assess completeness in two steps. First, we measure the detection completeness, i.e., the completeness of source detection and extraction with \texttt{SourceExtractor}. Then, we measure the selection completeness, i.e., the completeness of our selection cuts, as a function of the redshift, and the rest-UV- and noise properties of our sample galaxies.

\subsubsection{Detection Completeness}
\label{sec:detection_completeness}

We use the GaLAxy survey Completeness AlgoRithm 2 (\texttt{GLACiAR2}) software \citep{Carrasco2018, Leethochawalit2022} to perform injection-recovery simulations. Specifically, we inject model galaxies into the real NIRCam images, run \texttt{SourceExtractor}, and measure the fraction of recovered galaxies as a function of their input and output magnitude. Given the large size of the EGS, UDS, COSMOS, GOODS-S and GOODS-N fields, we run \texttt{GLACiAR2} on $7.5\arcmin\times7.5\arcmin$ cutouts of each image. The cutouts however cover a substantial part of each mosaic, including deeper and shallower parts as well as regions near the edge. We only include the F277W, F356W and F444W mosaics in the simulations, and an inverse-variance weighted stack of the three filters as the detection image.

For each field, we perform 10 iterations with 500 galaxies per iteration. The input galaxies have flat SEDs in $f_\nu$ ($\beta=-2$), are fixed to be at a redshift of $z=10$, and are injected in 42 magnitude bins ranging from 22.5 to 32.5 AB. An equal number of galaxies is injected in each magnitude bin, but their values of M$_{\rm UV}$ are sampled from a Schechter function \citep{Schechter1976} within the bin.

The model galaxies are assumed to have S\'ersic light profiles with 25\% of the galaxies having S\'ersic indices of 1 and 2 respectively, and the remaining 50\% having a S\'ersic index of 1.5. Their sizes follow a log-normal distribution centered on R$_{\rm eff}=0.36\,$kpc. Finally, we use 5 equally sized bins in inclination and eccentricity.

Each simulation yields a matrix specifying the number of recovered sources in each bin of input and output magnitude. The detection completeness in a bin of output magnitude is then given by a weighted mean over the recovered fraction in each input magnitude bin from which galaxies scattered into that output magnitude bin.

\subsubsection{Selection Completeness}
\label{sec:selection_completeness}

To assess the completeness of our sample selection (Section \ref{sec:sample_selection}), we roughly follow the methodology outlined in \citet{Harikane2022}. For each galaxy in our sample, we create mock SEDs, based on the 500 sets of values of M$_{\rm UV}$, $\beta$, and $z$, sampled \textit{together} from the \texttt{bagpipes} posterior to retain covariance information, and assuming a simple power law shape with a slope $\beta$. We apply IGM attenuation to the mock SEDs using the model from \citet{Inoue2014}, and convolve them with the corresponding NIRCam filter throughput curves to obtain synthetic fluxes in F115W, F150W, F200W, F277W, F356W, and F444W. Then, for each of the 500 sets of synthetic fluxes per source, we create 1000 realizations of the mock photometry, adding Gaussian noise to the flux in each filter, according to the 1$\sigma$ uncertainties for that source in our photometric catalog. Of those 1000 realizations, we first determine the fraction that would be detected in our \texttt{SourceExtractor} run by combining the fluxes and uncertainties in F277W, F356W, and F444W to estimate the SNR in the detection stack, and comparing this to the detection threshold we use in \texttt{SourceExtractor}. We then apply our selection cuts to the ``detected'' mock galaxies to get an estimate of the selection completeness. This way, we obtain 500 values of the selection completeness for each source, corresponding to the 500 draws of M$_{\rm UV}$, $\beta$, and $z$ from the \texttt{bagpipes} posterior. Since these completeness values are relatively uncertain, and low values of completeness significantly affect the inferred UVLFs, we cap them at 0.02 to avoid having single objects be weighted by $>50\times$. We discuss this further in Section \ref{sec:selection_function}.

We can then treat the two completeness values independently and obtain a total completeness by multiplying the detection completeness and the selection completeness for each of our sample galaxies, and at each draw from the \texttt{bagpipes} posterior. Mathematically, this can be expressed as,

\begin{equation}
\label{eq:compl_tot}
C_{\rm tot}(f_{\rm det}, {\rm M_{UV}}, \beta, z) = C_{\rm det}(f_{\rm det})\times C_{\rm sel}({\rm M_{UV}}, \beta, z)
\end{equation}

\noindent where $f_{\rm det}$ is the flux in the detection image, and

\begin{equation}
\label{eq:compl_sel}
C_{\rm sel}({\rm M_{UV}}, \beta, z) = P({\rm sel}\,|\,{\rm det}) = \frac{N_{\rm sel\,\land\,det}({\rm M_{UV}}, \beta, z)}{N_{\rm det}({\rm M_{UV}}, \beta, z)}
\end{equation}

\noindent where among the 1000 realizations $N_{\rm det}$ is the number of mock galaxies that would be detected by \texttt{SourceExtractor}, and $N_{\rm sel\,\land\,det}$ is the number that would be detected and also pass our selection cuts for a given dropout selection.

\subsection{Binned UV Luminosity Function}
\label{sec:binned_luminosity_function}

Measuring the UVLF means counting galaxies in bins of M$_{\rm UV}$ and redshift. In our case, the redshift bins are given by the redshift selection function of each of our dropout samples, characterized as described in the previous Section. 
We use bins of 1 magnitude in M$_{\rm UV}$, ranging from -22.5 to -18.5 for the F115W dropouts, and replace the three brighter bins by two bins of 1.5 magnitude for the F150W, and the F200W dropouts to account for the smaller number counts in these samples. In each bin, the number density of galaxies is given by
\begin{multline}
\label{eq:uvlf}
\Phi({\rm M_{UV,\,bin}},\,z_{\rm bin}) = \sum_{i=0,...,N} \frac{1}{V_{\rm eff,\,i}} = \\
\sum_{i=0,...,N} \frac{1}{C_{\rm tot, i}(f_{\rm det,\,i}, {\rm M_{UV,\,i}}, \beta_{\rm i}, z_{\rm i}) V({\rm M_{UV,\,bin}})}
\end{multline}

where the index i runs over all N sources in the bin with median UV-magnitude of ${\rm M_{UV,\,bin}}$ and median redshift of $z_{\rm bin}$. $V_{\rm eff,\,i}$ is the effective volume over which galaxy i can be observed and this can be expressed as the total completeness measured for source i according to equation \ref{eq:compl_tot} times the total survey volume $V$ at ${\rm M_{UV,\,bin}}$ which we compute as explained in Section \ref{sec:survey_area}.

\subsection{Uncertainties in the UVLF}
\label{sec:uncertainties}

In order to estimate the total uncertainty in the number density of galaxies in each bin of M$_{\rm UV}$, we need to take into account three independent contributions to the uncertainty: Poisson noise, the effect of measurement uncertainties in M$_{\rm UV}$, $\beta$, and $z$, and cosmic variance.

\subsubsection{Poisson Noise}
\label{sec:poisson_noise}

To estimate the Poisson uncertainty on the number count in a given M$_{\rm UV}$ bin, we use the frequentist central confidence interval (see \citealt{Maxwell2011}). In bins where we count 0 galaxies, we replace $N$ in Equation \ref{eq:uvlf} by $N_{\rm ulim}=3.8$, the 2$\sigma$ upper limit for a single-sided Poisson distribution as quoted in Table 1 of \citet{Gehrels1986}.

\subsubsection{Propagation of Uncertainties in M$_{\rm UV}$, $\beta$, and $z$}
\label{sec:propagation_of_uncertainties_in_muv}

We make use of the Bayesian nature of \texttt{bagpipes} and sample 500 SEDs from the posterior distribution from which we measure M$_{\rm UV}$, $\beta$, and $z$ as explained in Section \ref{sec:sed_fitting}. This allows us to then generate 500 realizations of the UVLF, applying Equation \ref{eq:uvlf} to each of the 500 values of z, M$_{\rm UV}$, and $\beta$ for each object. The median number densities among these 500 measurements in each M$_{\rm UV}$ bin yield our final measurement of the UVLF $\Phi$, and the scatter among them quantifies the effect of uncertainties in the photometry, the SED-fitting, and the completeness correction on $\Phi$.

\subsubsection{Cosmic Variance}
\label{sec:cosmic_variance}

Finally, despite the 35 independent lines of sight used in this work, our measurements are subject to cosmic variance. To estimate its contribution to the uncertainty of our UVLFs, we calibrate the cosmic variance to the \texttt{UniverseMachine} simulation suite \citep{Behroozi2019}, following \citet{Jespersen2025}. This directly incorporates scatter in the stellar-to-halo mass relation as well as possible effects from assembly and environmental biases \citep{Jespersen2022, Wu2024, Chuang2024}. We calibrate the cosmic variance in the same bins of M$_{\rm UV}$ in which we compute the UVLF as described above, and assuming the $M_*$-M$_{\rm UV}$ relation from \citet{Song2016}. Each field is approximated by a rectangle whose area matches the total area of the field.

To combine the cosmic variance in each magnitude bin across the different fields, we follow the procedure of \citet{Valentino2023}, 

\begin{equation}
\label{eq:cosmic_variance}
\sigma_{\mathrm{CV},\,\text{total}}=\sqrt{\frac{1}{\sum_{\text{fields }} \sigma_{\mathrm{CV},\,\text{field}}^{-2}}}
\end{equation}

\noindent where $\sigma_{\rm CV,\,field}$ is the fractional field-to-field variance in the number count of galaxies. Equation \ref{eq:cosmic_variance} assumes that all fields are independent. Since some PANORAMIC pointings overlap with legacy fields, we add their area to that of the respective legacy field instead of treating them separately. In each bin of M$_{\rm UV}$, we only include fields that are deep enough to contribute sources to that bin, i.e., whose median limiting magnitude (see Section \ref{sec:survey_area}) is higher than the central value of the M$_{\rm UV}$ bin respectively.

The resulting values are included as a contribution to the error bars on our UVLFs by combining the three sources of uncertainty in quadrature as,

\begin{equation}
\label{eq:uncertainties}
\sigma_{\Phi}^2 = \sigma_{\rm Poisson}^2 + \sigma_{\rm phot}^2 + \sigma_{\rm CV,\,\text{total}}^2\langle N\rangle^2
\end{equation}

\noindent where $\sigma_{\rm Poisson}$ is the Poisson uncertainty, $\sigma_{\rm phot}$ is the contribution from uncertainties in the photometry and the SED-fitting, and $\langle N\rangle$ is the completeness-corrected number count in a given bin of $z$ and M$_{\rm UV}$. 

\subsection{Survey Area and M$_{\rm UV}$ Limits}
\label{sec:survey_area}

To account for the varying depth across and between fields used in this work, we measure the survey area as a function of the M$_{\rm UV}$ limit up to which we can detect galaxies for each dropout sample. The M$_{\rm UV}$ limit is given by a combination of the SNR cut in the filter redward of the Lyman break in our selection (SNR(FXXXW$)>8$), and the requirement of a color $>1.5$ between that filter and the adjacent dropout filter.
We start from the weight (``wht'') maps from the DJA, which we combine with the exposure time maps to obtain ``full'' weight maps, including Poisson noise\footnote{\url{https://dawn-cph.github.io/dja/blog/2023/07/18/image-data-products/}}. Then, we convert those ``full'' weight maps into rms maps as ${\rm rms} = 1/\sqrt {\rm wht}$, and multiply each rms map by a scaling factor that is derived from placing apertures on empty parts of the respective flux map to account for correlated noise (see also \citealt{Weibel2024}). We then smooth the rms maps with a median filter with a box size of $7\times7$ pixels, roughly corresponding to the area of the circular aperture (radius of 4 pixels) that we use for our photometric catalogs. On the median-filtered rms map, we derive an $M_{\rm UV}$ limit for each pixel i as,

\begin{equation}
{\rm M_{UV,\,lim,\,i}} = 28.9 - 2.5\,{\rm log}(f_{\rm lim,\,i}) - d + 2.5\,{\rm log}(1 + z_{\rm bin})
\end{equation}

\noindent where $z_{\rm bin}$ is the median redshift of the respective bin, d is the distance modulus at $z_{\rm bin}$, and $f_{\rm lim,\,i}$ is defined as,

\begin{equation}
f_{\rm lim,\,i} = {\rm rms}_{\rm i} \cdot \sqrt{A_{\rm aper}}\cdot {\rm SNR}_{\rm thresh}
\end{equation}

\noindent representing the limiting flux at pixel i where rms$_{\rm i}$ is the rms value at pixel i,  and $A_{\rm aper}$ is the area of the circular aperture used to measure the photometry in pixels ($A_{\rm aper}=4^2\pi\approx50.3$).
For each redshift bin, we compute two $M_{\rm UV}$ limit maps: one in the dropout filter, with SNR$_{\rm thresh}=2$, and one in the filter redward of the break with SNR$_{\rm thresh}=8$. The first map specifies the $M_{\rm UV}$ values corresponding to the $2\sigma$ upper limit we apply when measuring the Lyman break color. To get the final map of $M_{\rm UV}$ limits, at each pixel, we add 1.5 to the value in the first map, compare it to the value at the same pixel in the second map, and use the higher value of the two. This is because in order to select a galaxy, the filter redward of the break has to be deep enough to \textit{both} detect the galaxy at ${\rm SNR}>8$, and measure a color $>1.5$ relative to the $2\sigma$ upper limit in the dropout filter.

We then count the number of pixels whose value of M$_{\rm UV,\,lim\,i}$ lies within small bins of M$_{\rm UV}$ with bin edges given by M$_{\rm UV,\,1}$, ..., M$_{\rm UV,\,j}$, M$_{\rm UV,\,j+1}$, ... and bin sizes of 0.05 mag. 
For an M$_{\rm UV}$ bin used when computing the UVLF with bin edges M$_{\rm UV,\,min}$ and M$_{\rm UV,\,max}$, we then obtain the area over which we select galaxies in that bin as,

\begin{equation}
\label{eq:survey_area}
{\rm A}_{\rm bin} = \sum_{{\rm j\,:\, M_{UV,\,j}>M_{UV,\,min}}}g({\rm M_{UV,\,j}})\sum_i f_{\rm i}({\rm M_{UV,\,j}})
\end{equation}

\noindent where

\begin{equation}
\label{eq:sa_sub1}
f_{\rm i}({\rm M_{UV,\,j}}) = 
\begin{cases} 
1 & {\rm if}\,\,\,{\rm M_{UV,\,j-1}<{\rm M_{UV,\,lim,\,i}} < {\rm M_{UV,\,j}}} \\
0 & {\rm else}
\end{cases}
\end{equation}

\noindent determines if pixel i lies in the small 0.05 mag M$_{\rm UV}$ bin j, and

\begin{equation}
\label{eq:sa_sub2}
g({\rm M_{UV,\,j})} = 
\begin{cases} 
\frac{\rm M_{UV,\,j} - M_{UV,\,min}}{\rm M_{UV,\,max} - M_{UV,\,min}} & {\rm if}\,\,\,{\rm M_{UV,\,j} < M_{UV,\,max}}\\
1 & {\rm else}
\end{cases}
\end{equation}

\noindent measures the fraction of the UVLF bin (M$_{\rm UV,\,min}$, M$_{\rm UV,\,max}$) between M$_{\rm UV,\,min}$ and the upper edge M$_{\rm UV,\,j}$ of the smaller bin (M$_{\rm UV,\,j-1}$, M$_{\rm UV,\,j}$).

We note that in practice, a galaxy in our sample with e.g., a measured M$_{\rm UV}=-19$ is not guaranteed to be observed in a part of the mosaic with a nominal M$_{\rm UV}$ limit $>-19$ due to the noise in the data that can up-scatter fainter galaxies above our selection threshold. This is, however, accounted for in our completeness correction, and the area computed as described above provides a more accurate estimate of the true area over which galaxies are selected compared to approaches determining an M$_{\rm UV}$ limit per field.

We slightly simplify this procedure for the Abell-2744 cluster field where we measure the median M$_{\rm UV}$ limit and then assume this to be constant across the field. Using the strong lensing model described in Section \ref{sec:lensing}, we can then compute the area above a given M$_{\rm UV}$ limit as the area with a magnification larger than what is required to reach the respective depth. Note that this is done directly in the source plane in order to obtain the volume effectively probed with the available observations.

\subsection{Fitting the UVLF}
\label{sec:fitting}

We fit two parametric curves to the obtained UVLFs, a single Schechter function, and a double power-law (DPL).

These functions can be described as,

\begin{multline}
\label{eq:phi_schechter}
\Phi_{\rm Schechter}({\rm M_{UV}}) = 0.4\,{\rm ln(10)}\,\Phi^*10^{-0.4({\rm M_{UV}} - {\rm M_{UV}^*})(\alpha +1)}\times \\
{\rm exp(-10^{-0.4({\rm M_{UV}} - {\rm M_{UV}^*})})}
\end{multline}

\noindent and 

\begin{equation}
\label{eq:phi_dpl}
\Phi_{\rm DPL} = \frac{\Phi^*}{10^{-0.4({\rm M_{UV}} - {\rm M_{UV}^*})(\alpha +1)} + 10^{-0.4({\rm M_{UV}} - {\rm M_{UV}^*})(\beta +1)}}
\end{equation}

\noindent To account for Eddington bias introduced by the uncertainties in M$_{\rm UV}$, we convolve the fitting functions with a Gaussian before fitting, i.e., we fit the function

\begin{equation}
\label{eq:phi_conv}
\Phi_{\rm conv}({\rm M_{UV}}) = \int^\infty_{-\infty}\Phi_{\rm param}(x)\mathrm{G}({\rm M_{UV}} - x)dx
\end{equation}

\noindent to the observed UVLF where G is a Gaussian whose width is given by the median 1$\sigma$ uncertainty in M$_{\rm UV}$ in each dropout sample, and $\Phi_{\rm param}$ stands for $\Phi_{\rm Schechter}$ and $\Phi_{\rm DPL}$ respectively.

We perform the fitting to each of the 500 UVLFs sampled from the \texttt{bagpipes} posterior, and use the median, as well as the 16th and 84th percentiles of the derived fitting parameters as our final estimates and their uncertainties.

Since we cannot fit for all the parameters simultaneously at $z\sim13$ and $z\sim17$ (see Section \ref{sec:fitting_results}), we also adopt the DPL parametrization from \citet{Donnan2024} to determine the values of ${\rm M_{UV}^*}$, $\alpha$ and $\beta$ at the median redshift of our samples respectively, and only fit for the normalization $\Phi^*$.

\section{Results}
\label{sec:results}

\subsection{High-Redshift Candidates}
\label{sec:high_redshift_candidates}

Subsequently, we discuss our three samples, mentioning some sources that have already been published, as well as highlighting the most compelling new candidates identified in this work.

\subsubsection{$z\sim10$ Sample}
\label{sec:z10_sample}

Spectroscopically confirmed sources in our $z\sim10$ sample include two sources in the EGS from \citet{ArrabalHaro2023}, eight sources in the A2744 cluster field (four from \citealt{Fujimoto2024} and four from \citealt{Napolitano2025}), as well as GN-z11 \citep{Oesch2016, Bunker2023}. References or program IDs for all spectra of our $z\sim10$ candidates are provided in Table \ref{tab:z10_candidates}. We further select two galaxies that were first mentioned as $z\sim9-10$ candidates (in addition to GN-z11) in \citet{Oesch2014} namely their GN-z10-2, and GN-z9-1.

\begin{figure*}
    \begin{center}
        \includegraphics[width=0.87\columnwidth]{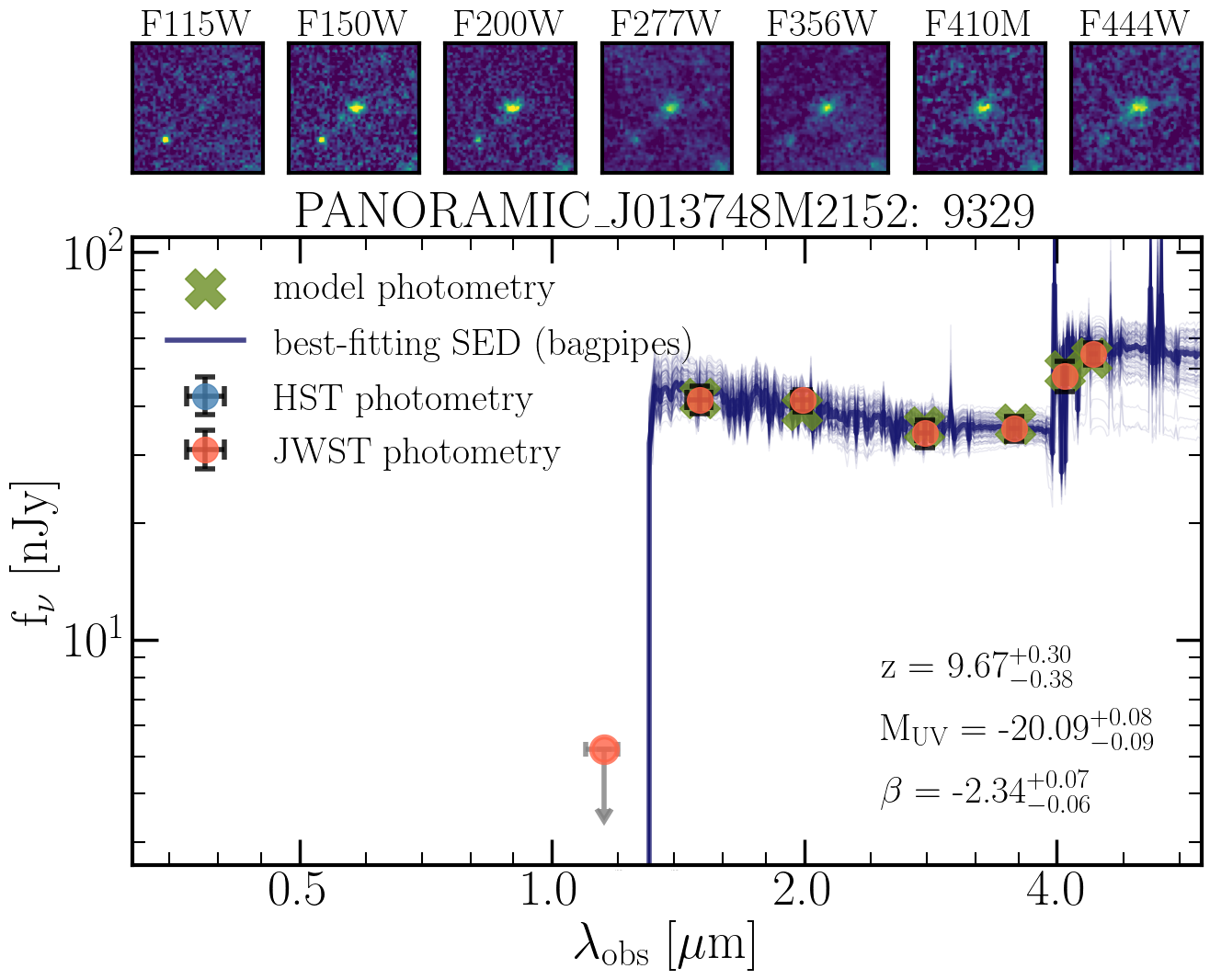}
        \includegraphics[width=0.87\columnwidth]{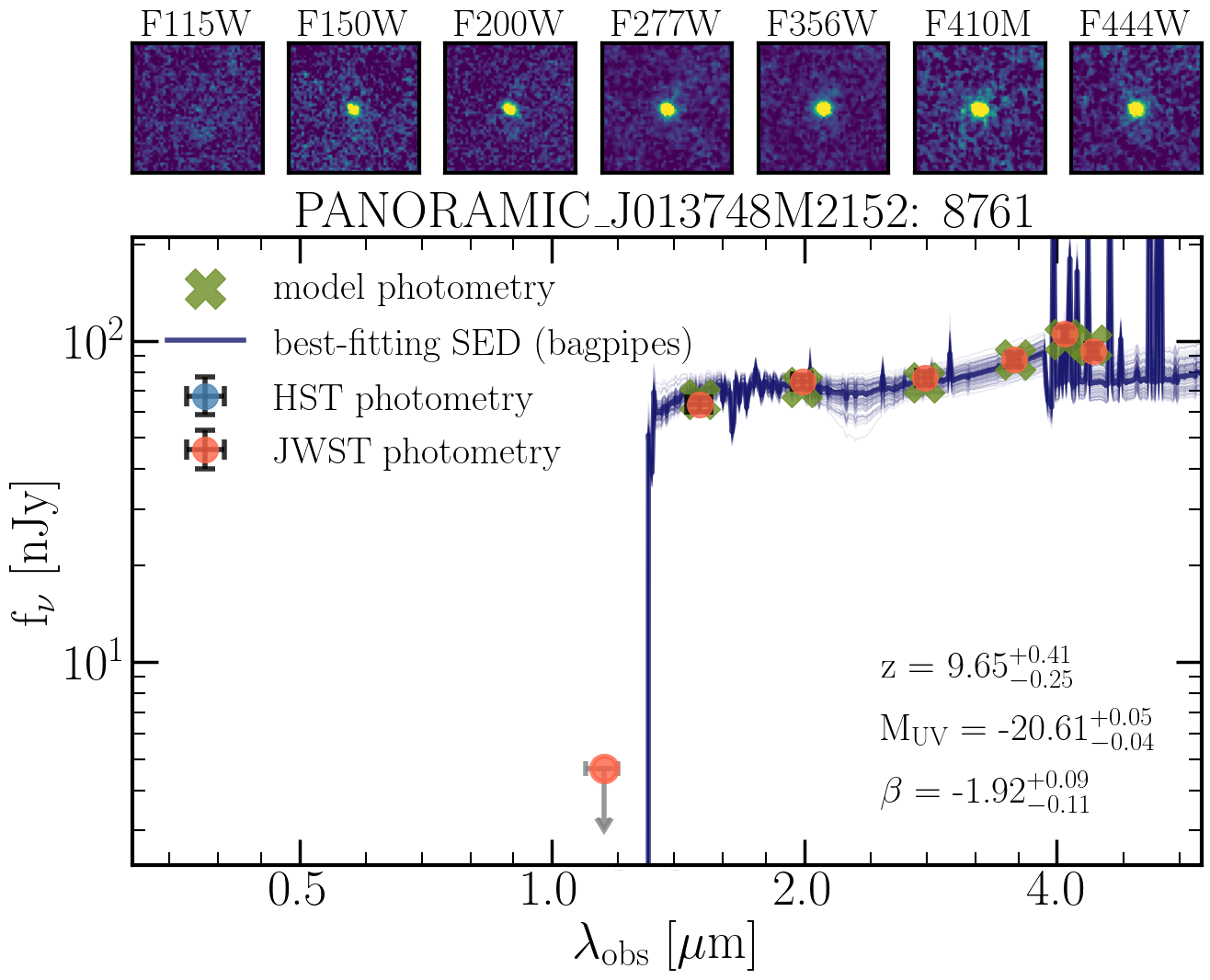}
        
    \end{center}
    \caption{Two convincing $z\sim10$ candidates identified as F115W dropouts in PANORAMIC. Both are found in the same pointing outside of legacy fields, and they are fit at very similar photometric redshifts. Assuming they indeed lie at the same redshift, their physical separation is only $82\,$kpc. The thin blue lines show 100 samples of the best-fitting SED from the \texttt{bagpipes} posterior respectively.}
    \label{fig:z10_candidates}
\end{figure*}

In Figure \ref{fig:z10_candidates}, we highlight two of the most striking candidates identified in PANORAMIC imaging outside of any legacy fields. They are both located in the same pointing (J013748M2152), and their photometric redshifts are consistent with each other, suggesting a physical separation of only 82 proper kpc. ID 9329 shows hints of a Balmer break as traced by F410M and F444W. While difficult to disentangle from line emission, a tentative Balmer break is observed in $>10$\% of all galaxies in the F115W dropout sample. If confirmed, this suggests that bursty star formation is common at $z\sim10$, even at relatively high masses (up to $M_*\sim10^9\,{\rm M_\odot}$ as indicated by our \texttt{bagpipes} fits).

\subsubsection{$z\sim13$ Sample}
\label{sec:z13_sample}

Our $z\sim13$ sample contains 4 objects with spectroscopic redshifts: GHZ2/GLASS-z12 \citep{Castellano2024}, GS-z12-0 from \citet{CurtisLake2023}, JADES-GS-z14-0 from \citet{Carniani2024}, and PAN-z14-1 from \citet{Donnan2026} (see Table \ref{tab:z13_candidates}).

\begin{figure*}
     \begin{center}
     \includegraphics[width=0.87\columnwidth]{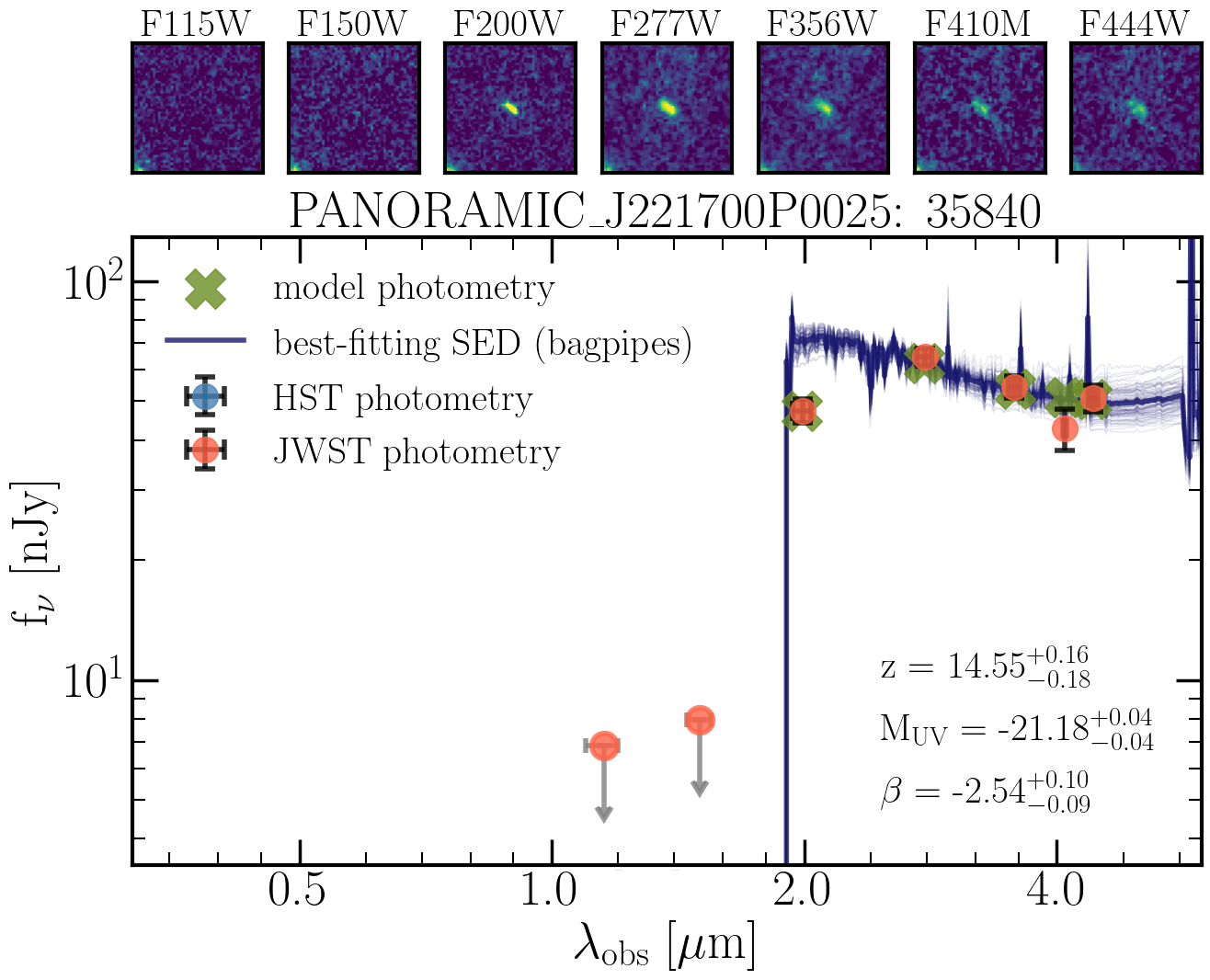}
     \includegraphics[width=0.87\columnwidth]{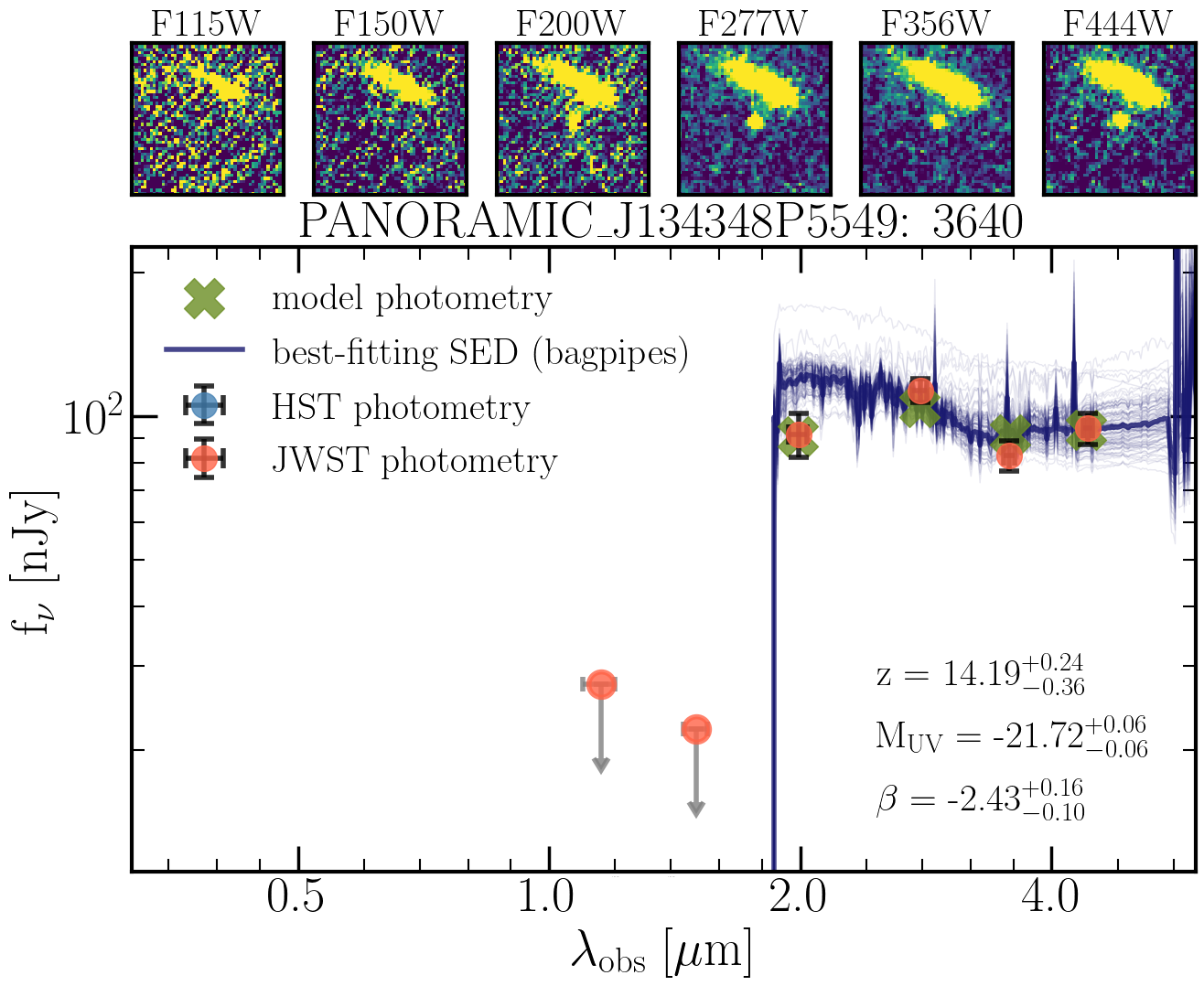}
     \end{center}
     \caption{Same as Figure \ref{fig:z10_candidates} with the most promising $z\gtrsim13$ candidates from PANORAMIC, identified as F150W dropouts. The source on the left has been recently confirmed to lie at $z_{\rm spec}=13.53$ \citep{Donnan2026}. ID 3640 on the right shows a similar photometric redshift, but, if confirmed, is even more luminous at M$_{\rm UV}\sim-21.7$.}
     \label{fig:z13_candidates}
 \end{figure*}

We show two promising $z\sim13-14$ candidates identified from PANORAMIC imaging in Figure \ref{fig:z13_candidates}. ID 35840, initially discovered by McLeod et al. (in prep.), and now confirmed to lie at $z_{\rm spec}=13.53$ in \citet{Donnan2026}, shows a blue UV-continuum, and a clear break between F200W and F150W (a factor of 6, or 2 mag, with a 2$\sigma$ upper limit in F150W). The slight drop in F200W indicates that the Lyman break is inside that filter, setting the photometric redshift to $z\sim14.5$. However, as discussed in \citet{Donnan2026}, this drop likely stems from a combination of noise and a Lyman-$\alpha$ damping wing, causing us to over-estimate the photometric redshift of this source by $\Delta z\sim1$.
The source with ID 3640 on the right is identified in significantly shallower imaging (5$\sigma$ depth in F200W of 27.8 mag, compared to 29.3 mag in the J221700P0025 pointing in which ID 35840 is found), but nevertheless shows a break strength of 1.54 mag between F200W and F150W because it is even more luminous at M$_{\rm UV}\sim-21.7$ making it the most luminous galaxy in our F150W dropout sample.

Another one of our $z\sim13$ candidates, identified from a combination of CEERS and BEACON imaging, has been highlighted as a robust high-z candidate in \citet{Zhang2026}. However, as mentioned in \citet{Donnan2026}, if only CEERS-imaging is considered, this source remains undetected, and it only appears in the BEACON imaging that was taken later. Given the brightness of the source (supposed M$_{\rm UV}\sim-20.3$ at $z\sim13.6$), this means that it is a transient rather than a high redshift galaxy. We remove this source from our sample which leaves us with 10 F150W dropouts.

\vfill\null

\subsubsection{$z\sim17$ Sample}
\label{sec:z17_sample}
\vspace{-0.1cm}

We discuss the four F200W dropouts that passed our selection cuts below and conclude that none of them is a plausible $z\sim17$ candidate. First, we select the interloper known as the Schr\"odinger galaxy \citep{Naidu2022, Harikane2023, Donnan2023, ArrabalHaro2023b}. Then, we identify an extremely promising candidate in the A2744 cluster field with a plausible blue UV continuum and a sharp break between F277W and F200W, indicating a redshift of $\sim18$ at M$_{\rm UV}\sim-20.1$. However, as has been discussed in \citet{Castellano2025}, this source turns out to be a transient. It was completely absent from epoch 1 imaging by GLASS in June 2022, and only showed up in epoch 2 imaging in November 2022, where unfortunately it fell into the chip gap of the NIRCam/SW detector and thus created a fake SW dropout source when imaging from both epochs is combined.

\begin{figure}
     \centering
     \includegraphics[width=0.47\textwidth]{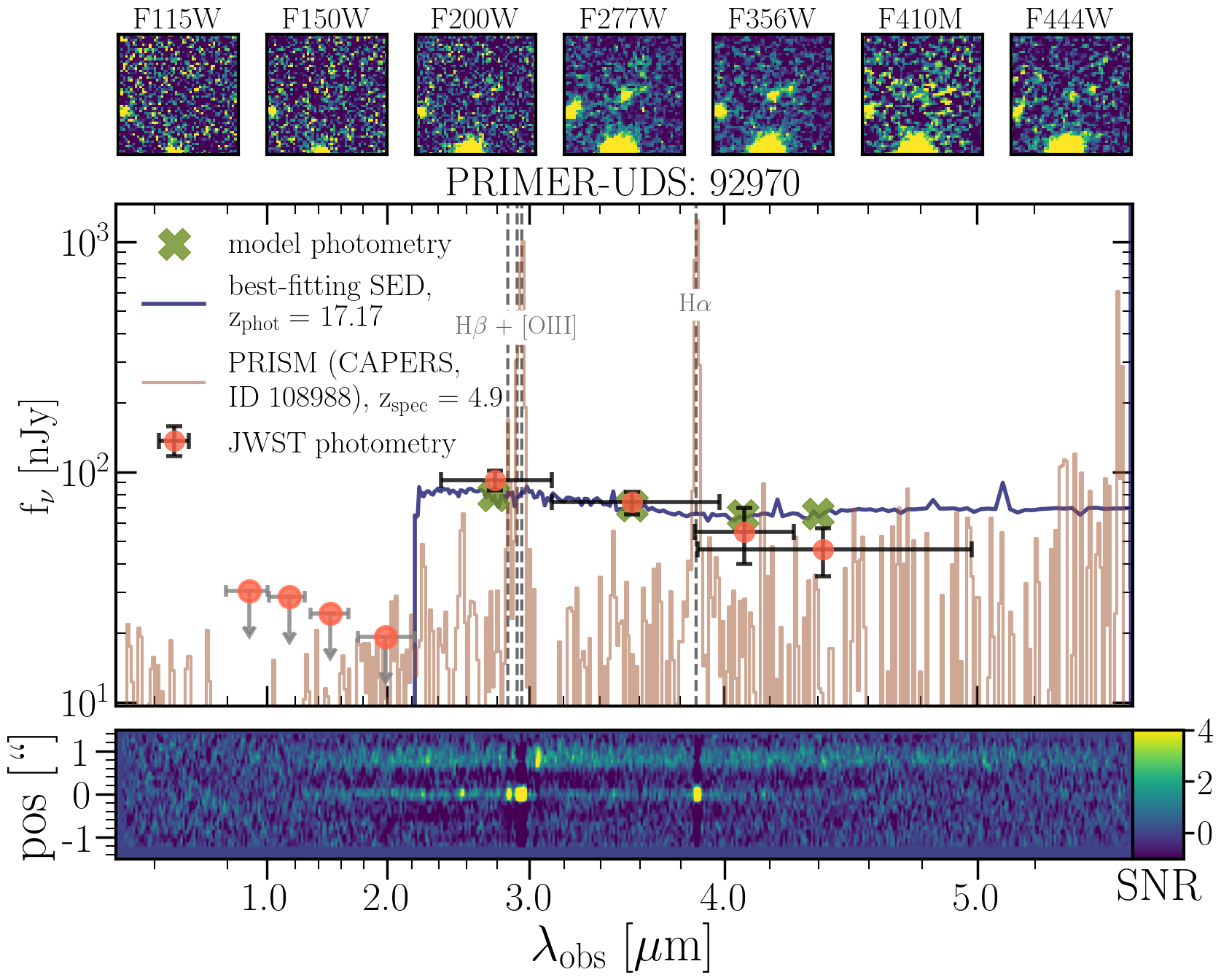}
     \caption{One of our F200W dropout candidates with its NIRSpec/PRISM spectrum from CAPERS revealing the [O\,{\sc iii}], H$\beta$, and H$\alpha$ emission lines which boost all four LW filters (F277W, F356W, F410M, and F444W) at $z_{\rm spec}=4.9$. The 1D spectrum is shown as the brown line, and the SNR of the 2D spectrum is shown in the bottom panel to better illustrate the clear emission line detections.}
     \label{fig:z17_capers}
 \end{figure}

The third $z\sim17$ candidate we select lies in the UDS field and is shown in Figure \ref{fig:z17_capers}. Its photometry, if it were at $z\sim17$, indicates a blue UV-continuum ($\beta=-2.29\pm0.26$). It lies next to a brighter source that also becomes fainter at shorter wavelengths probed by JWST/NIRCam, and that has a photometric redshift of $z_{\rm phot}=4.85\pm0.19$. Indeed, our candidate has a spectrum from CAPERS which reveals the [O\,{\sc iii}], H$\beta$, and H$\alpha$ lines lining up to boost the flux of all the LW filters, including F410M, and unambiguously putting the source at $z_{\rm spec}=4.9$, identical to the Schr\"odinger galaxy.

\begin{figure}
     \centering
     \includegraphics[width=0.4\textwidth]{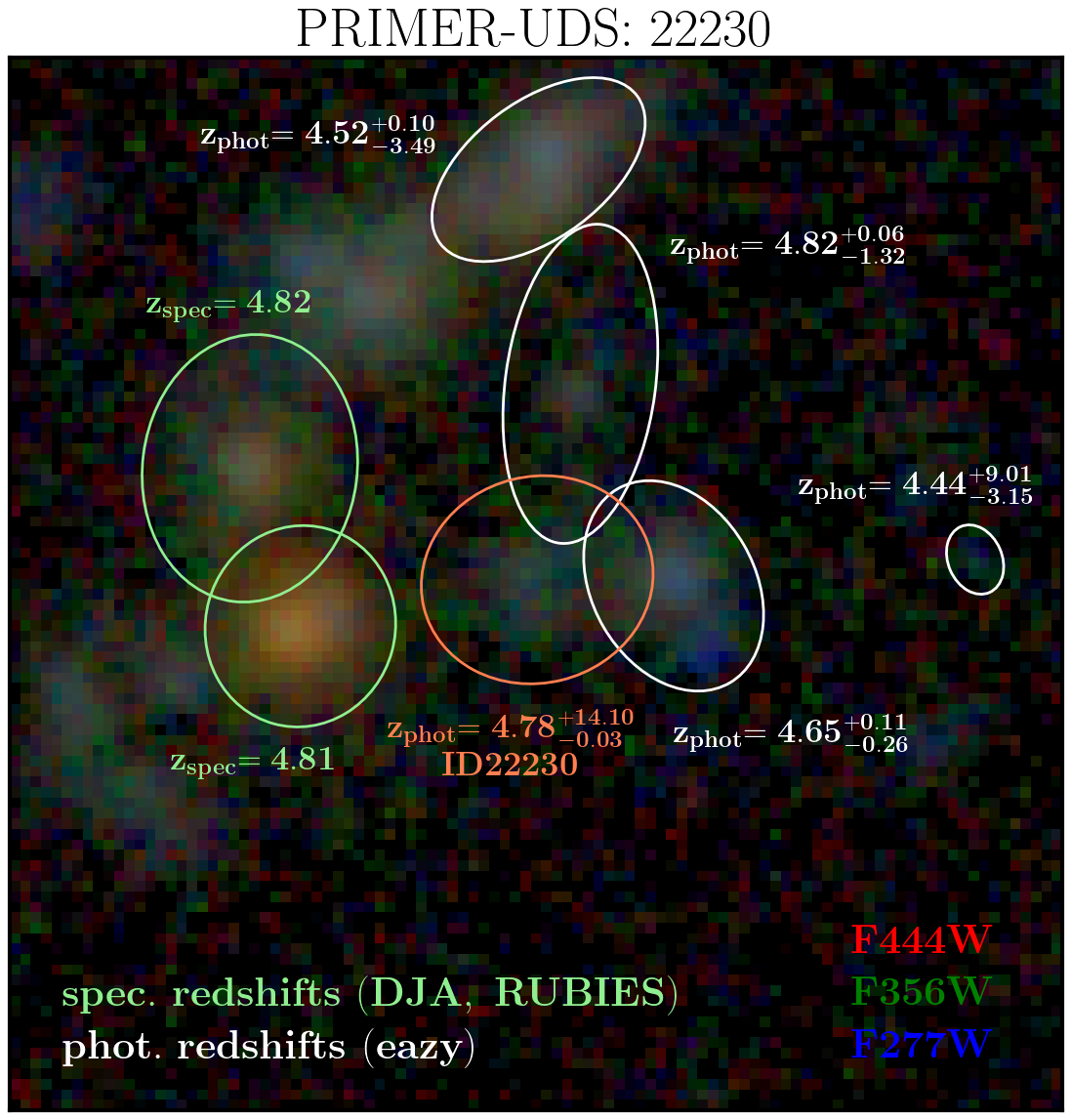}
     \caption{$4.04\arcsec\times4.04$\arcsec\ RGB imaging cutout of one of our F200W dropout $z\sim17$ candidates and its surroundings, highlighting sources with photometric and spectroscopic redshifts consistent with or close to $z=4.9$, where at least two interlopers contaminating $z\sim17$ F200W dropout searches are known.}
     \label{fig:z17_neighbors}
 \end{figure}

Our last F200W dropout candidate lies in a relatively crowded part of the UDS field. It shows an extended morphology and a low surface brightness which seems atypical of high redshift galaxies. More importantly, as many as four immediate neighbors of the source have photometric redshifts consistent within 1$\sigma$ with a redshift $<0.3$ different from the notorious $z\sim4.9$ which is also the preferred solution identified by \texttt{eazy} for the source itself.Furthermore, another two of its immediate neighbors have spectroscopic redshifts from RUBIES \citep{DeGraaff2025}, locating them at $z_{\rm spec}=4.81$ and $4.82$. We therefore argue that this source is likely not at $z\sim17$. An RGB imaging cutout of the source and its surroundings, with potential neighboring galaxies labeled with their photometric and spectroscopic redshifts is shown in Figure \ref{fig:z17_neighbors}.

When computing the UVLF below, we assume that we did not identify any plausible $z\sim17$ galaxies and only show upper limits on the number density at that redshift. 

\subsection{UV Luminosity Functions}
\label{sec:uvlf}

\begin{figure*}
    \begin{center}
        \includegraphics[width=0.97\columnwidth]{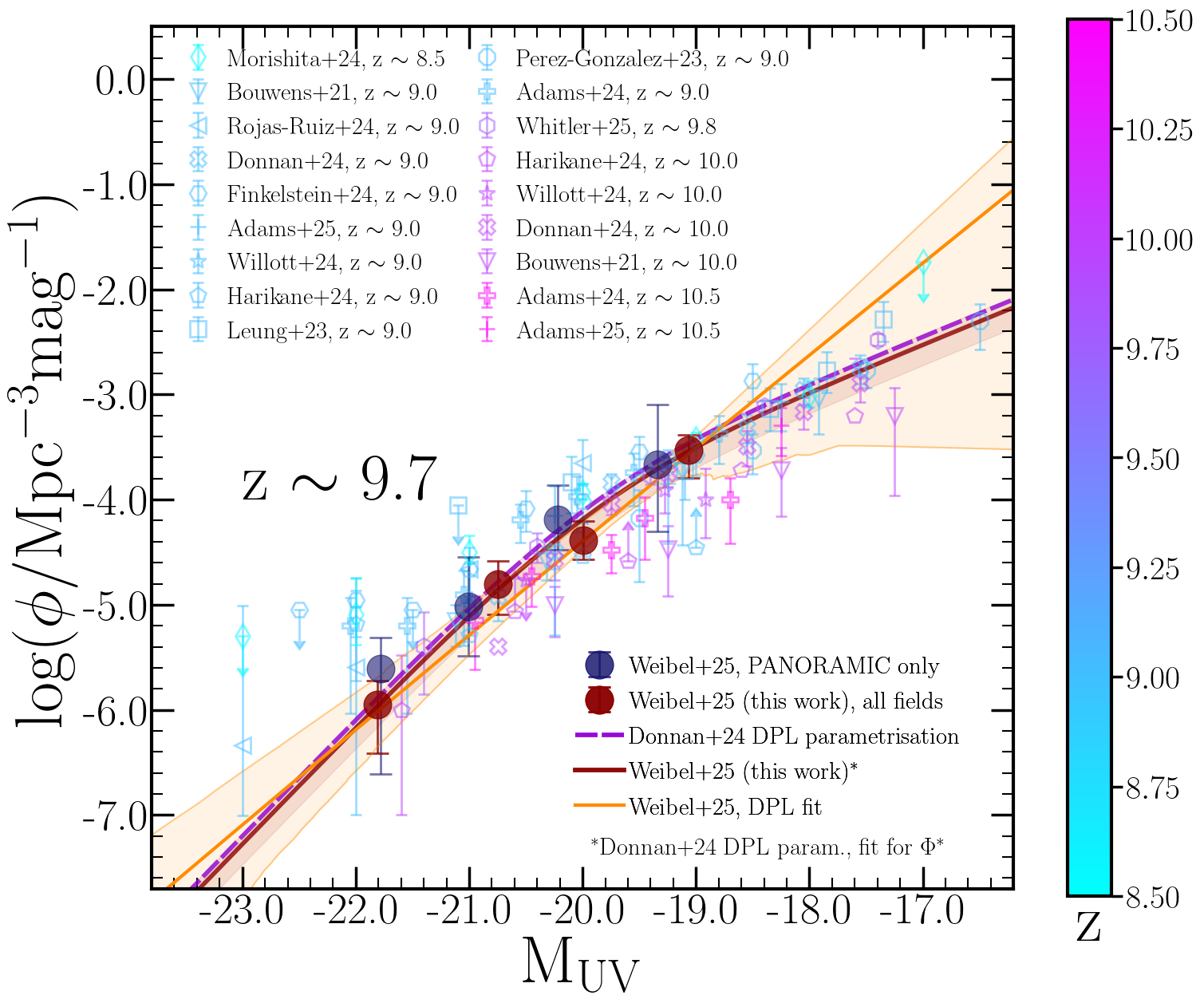}
        \includegraphics[width=0.86\columnwidth]{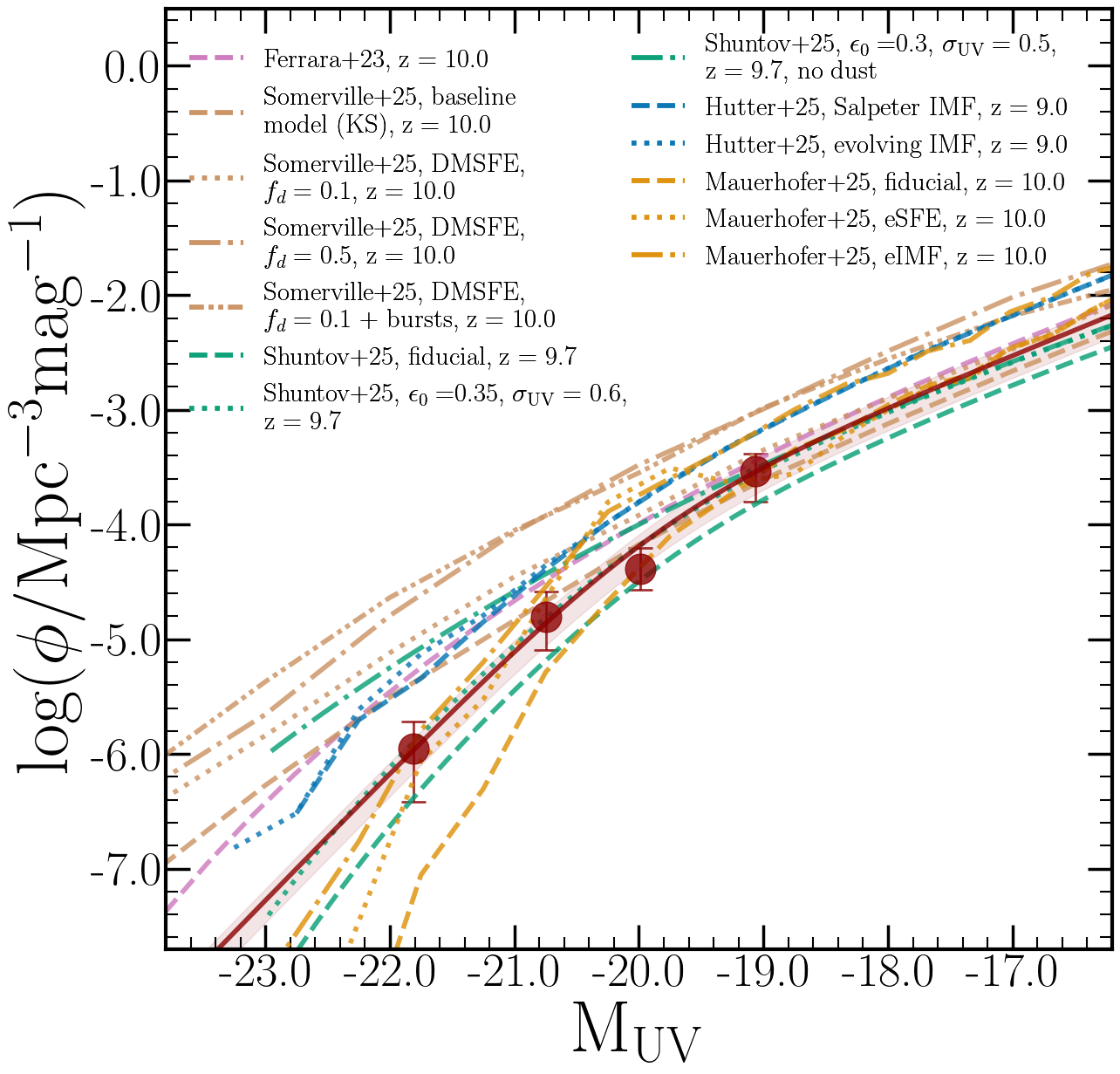}
        \centering
        \includegraphics[width=0.97\columnwidth]{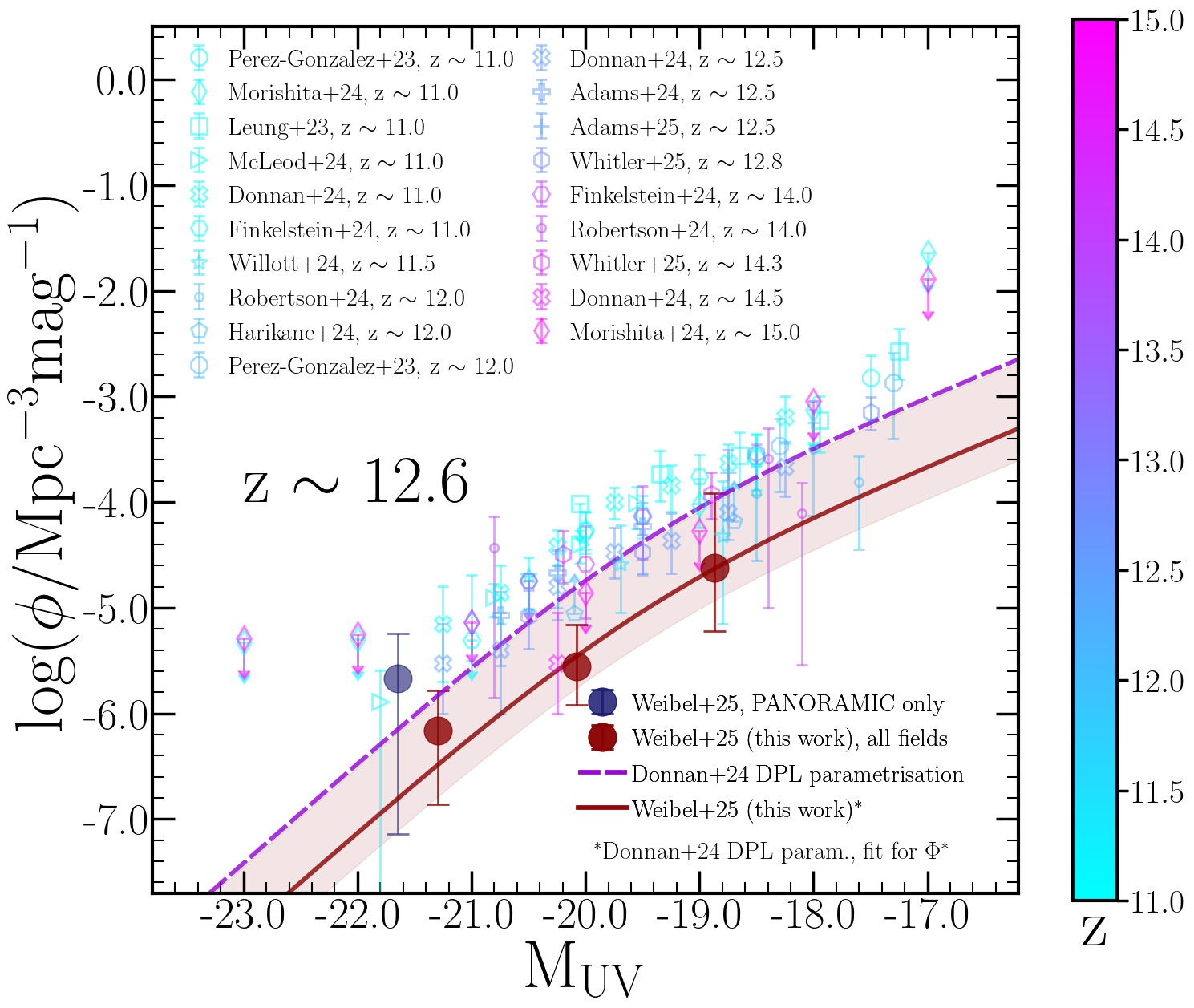}
        \includegraphics[width=0.86\columnwidth]{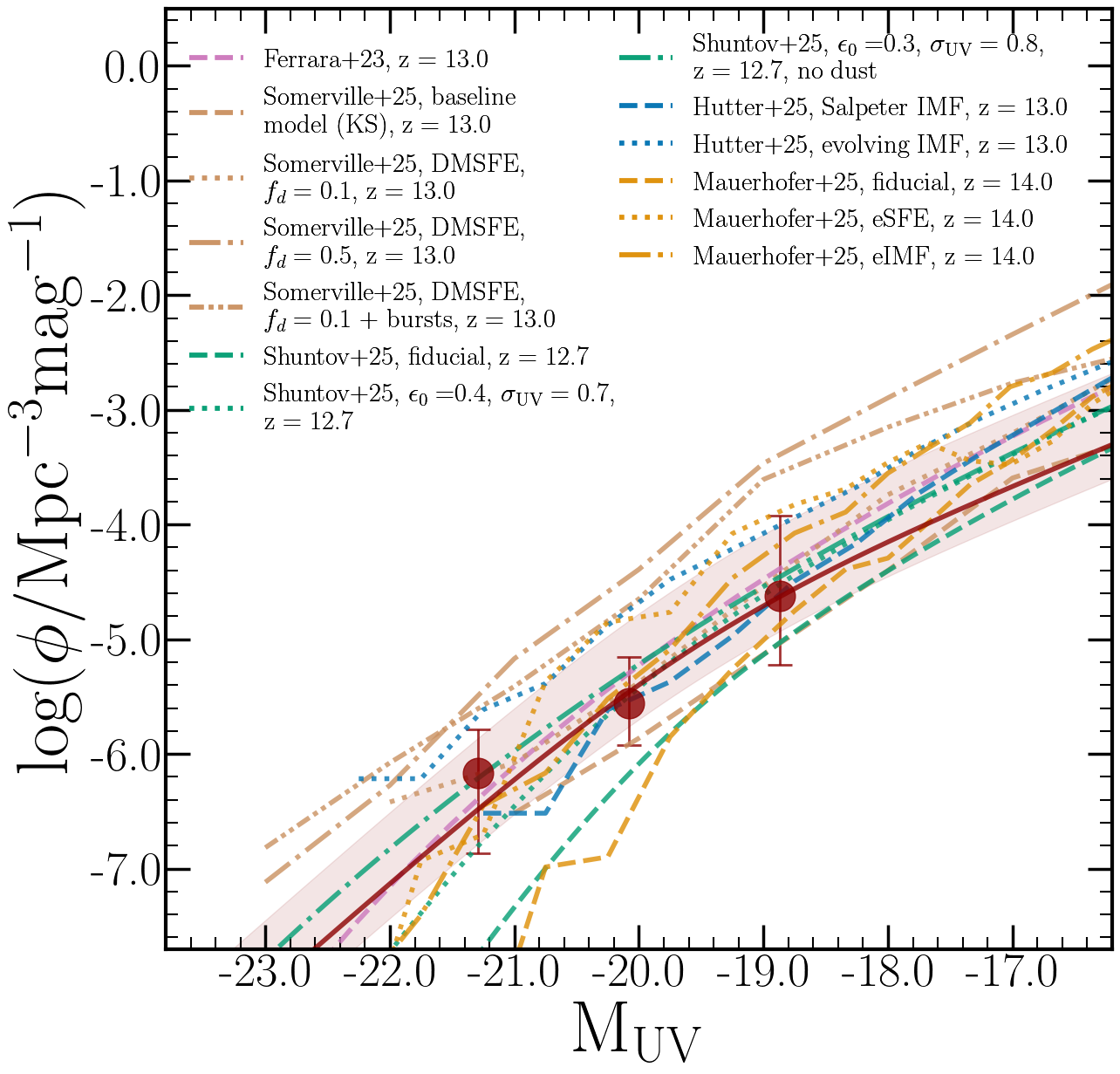}
        \centering
        \includegraphics[width=0.97\columnwidth]{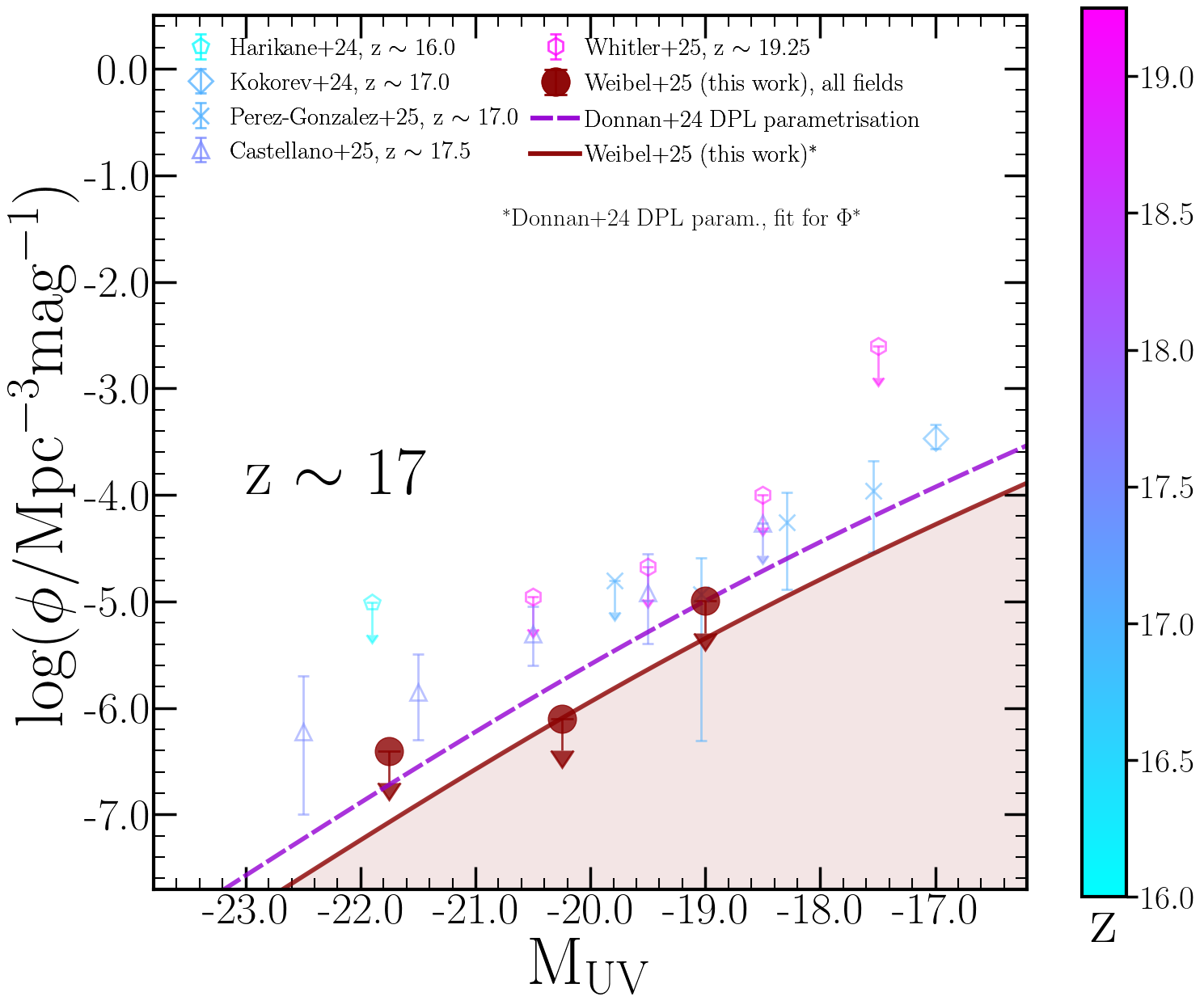}
        \includegraphics[width=0.86\columnwidth]{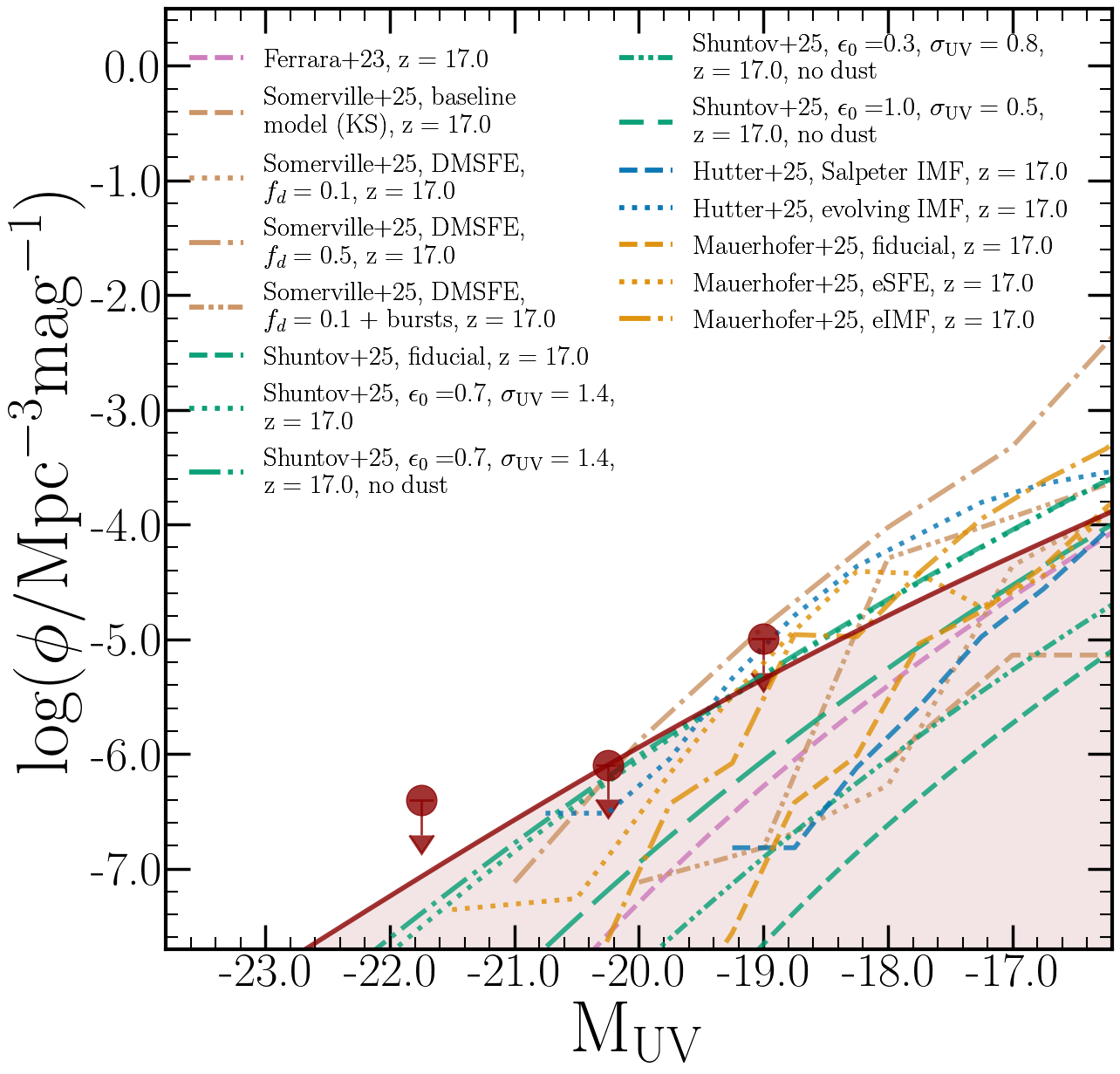}
    \end{center}
    \caption{Binned UVLFs inferred from our F115W (top), F150W (middle), and F200W (bottom) dropout samples. Number densities inferred from our full sample are shown as red dots, and those from PANORAMIC data only as blue dots. The red solid lines and shaded regions indicate our best-fitting DPL curve, where we adopted the parametrization from \citet{Donnan2024} (dark violet dashed lines in the left panels), and only fit for the normalization $\Phi^*$. In the top left panel (F115W dropouts) we further show our best-fitting DPL function where we fit for all parameters simultaneously. We compare to observational results on the left, and to models and simulations on the right, as indicated in the Figure legends and discussed in the text. Observational results are color-coded by their redshift within the redshift range covered by each panel.}
    \label{fig:uvlfs}
\end{figure*}

We present the UVLFs measured for our three dropout samples, corresponding to redshifts of $z\sim10$, $z\sim13$, and $z\sim17$ in Figure \ref{fig:uvlfs}. All our measured values and uncertainties are listed in Table \ref{tab:uvlf_values}. There are two panels for each dropout sample in Figure \ref{fig:uvlfs}, both showing our UVLF and best-fitting DPL curve (adopting the parametrization from \citet{Donnan2024}, and only fitting for $\Phi^*$) with uncertainties. The M$_{\rm UV}$ values at which the points are plotted are given by the median M$_{\rm UV}$ of the sample galaxies in a given bin. In the left panels, we compare to observational results from \citet{Bouwens2021}, \citet{Harikane2023}, \citet{PerezGonzalez2023b}, \citet{Leung2023}, \citet{McLeod2024}, \citet{Harikane2024}, \citet{Adams2024}, \citet{Willott2024}, \citet{Finkelstein2024}, \citet{Robertson2024}, \citet{Donnan2024}, \citet{Rojas-Ruiz2024}, \citet{Kokorev2025}, \citet{Morishita2025}, \citet{Whitler2025}, \citet{Adams2025}, \citet{PerezGonzalez2025}, and \citet{Castellano2025}.

The blue dots in the upper left and middle left panels represent number density estimates based on PANORAMIC data only. For this comparison, we only include the 30 PANORAMIC pointings outside of any legacy imaging. The measurements are plotted 0.2 magnitudes lower in $M_{\rm UV}$ for better visual separation. They are all consistent within error bars with the measurements including all fields, illustrating the power of pure parallel imaging in only 6 NIRCam filters to provide an unbiased view of the galaxy population at $z\gtrsim9$. In the F150W dropout sample, there are only two PANORAMIC candidates outside of any legacy fields which are both shown in Figure \ref{fig:z13_candidates}, and yield the number density shown as a blue point at the bright end of the UVLF in the middle left panel. This is consistent with, but slightly higher than the measurement from the full data set. As discussed in Section \ref{sec:z13_sample}, while the candidate with ID 35840 has been confirmed spectroscopically \citep{Donnan2026}, the supposedly even brighter candidate with ID 3640 is identified in shallower imaging and leaves more room for a low-z solution. If we remove this object from the sample, the resulting number density in the brightest bin at $z\sim12.6$ drops to log($\Phi/{\rm Mpc}^{-3}{\rm mag}^{-1})=-6.38_{-0.82}^{+0.27}$ for the full sample, and to log($\Phi/{\rm Mpc}^{-3}{\rm mag}^{-1})=-5.98_{-\infty}^{+0.31}$ for the PANORAMIC data only.

Splitting the uncertainties on our measured number densities in the three components specified in Equation \ref{eq:uncertainties}, we find that cosmic variance dominates the error in the brightest bin at $z\sim10$, followed by the Poisson uncertainty. In fainter bins, $\sigma_{\rm phot}$ starts to dominate due to the uncertainty in the completeness correction, followed by $\sigma_{\rm CV}$ while $\sigma_{\rm Poisson}$ becomes negligible. At $z\sim13$, the upper error bar (i.e., the difference between the 84th percentile and the median) is always dominated by $\sigma_{\rm phot}$, driven by uncertainties in the completeness correction. In the faintest bin, $\sigma_{\rm phot}$ also dominates the lower error bar, but in the two brighter bins, $\sigma_{\rm CV}$ followed by $\sigma_{\rm Poisson}$ have a stronger effect on the lower error bar.

From our F115W dropout sample at a median redshift of $z\sim9.7$, we measure a UVLF that is consistent with most literature results at comparable redshifts. Our measured number density in the brightest M$_{\rm UV}$ bin has more constraining power than other literature results thanks to the large area probed and the many independent lines of sight, and confirms the large abundance of bright galaxies at $z\sim10$, suggesting a shallow slope at the bright end of the UVLF out to M$_{\rm UV}\sim-22$ (see Section \ref{sec:fitting_results}).

At $z\sim12.6$, the median redshift of our F150W dropout sample, our measurements lie somewhat below published results. In the brightest bin, at M$_{\rm UV}\sim-21$, our data point is consistent within 1$\sigma$ with \citet{Donnan2024}, who measured the UVLF at $z\sim12.5$, while in the faintest bin (M$_{\rm UV}\sim-19$), our measurement is consistent with all displayed results due to the large uncertainty driven by $\sigma_{\rm phot}$ discussed above. At intermediate UV luminosities (M$_{\rm UV}\sim-20$), however, we find relatively few galaxies, resulting in $>1\sigma$ discrepancies with most literature results at similar redshifts, and also below the lower limit inferred by \citet{Harikane2024} from a pure spectroscopic sample at $z\sim12$. This then suggests a somewhat faster decrease in the overall number density of galaxies from $z\sim10 - 13$ compared to most literature results, as we will further explore below.

Moving to our highest redshift bin, our upper limits at $z\sim17$ are more constraining than previously published upper limits thanks to the larger survey area probed here. At M$_{\rm UV}\sim-19$, our upper limit coincides with the measurement of \citet{PerezGonzalez2025} and the parametrization from \citet{Donnan2024}. We do not probe faint enough magnitudes to directly compare to the fainter bins from \citet{PerezGonzalez2025}, or \citet{Kokorev2025}, but our fit to the upper limits, extrapolated to fainter magnitudes with the \citet{Donnan2024} parametrization suggests lower number densities than their measurements. As discussed in \citet{Kokorev2025}, if their number density is correct, a steep faint-end slope of the UVLF is required to match both their measurement and our upper limit at M$_{\rm UV}\sim-19$. At M$_{\rm UV}\sim-20$, our 2$\sigma$ upper limit lies slightly below the DPL parametrization by \citet{Donnan2024}, and the measurements from \citet{Castellano2025} who however also specify upper limits at roughly the same number densities as their measurements, reflecting the case where their candidates turn out to be low redshift interlopers.

\begin{table}
\centering
\caption{Measured UVLF-values from the three dropout samples, corresponding to $z\sim10$, 13 and 17.}
\label{tab:uvlf_values}
\begin{tabular}{ccc}
\hline
& & \\[\dimexpr-\normalbaselineskip+1pt]
Redshift bin & M$_{\rm UV}$ & log($\Phi/{\rm Mpc^{-3}\, mag^{-1}})$ \\
& & \\[\dimexpr-\normalbaselineskip+1pt]
\hline
\hline
\multirow{6}{*}{$z\sim10$} 
& & \\[\dimexpr-\normalbaselineskip+2pt]
& -21.81 & $-5.95^{+0.23}_{-0.46}$\\[5pt]
& -20.75 & $-4.80^{+0.22}_{-0.29}$\\[5pt]
& -19.99 & $-4.38^{+0.18}_{-0.18}$\\[5pt]
& -19.07 & $-3.53^{+0.15}_{-0.27}$\\[5pt]\hline
\multirow{4.545}{*}{$z\sim13$} 
& & \\[\dimexpr-\normalbaselineskip+2pt]
& -21.29 & $-6.17^{+0.38}_{-0.70}$\\[5pt]
& -20.08 & $-5.56^{+0.40}_{-0.37}$\\[5pt]
& -18.87 & $-4.63^{+0.71}_{-0.60}$\\[5pt]\hline
\multirow{4.545}{*}{$z\sim17$} 
& & \\[\dimexpr-\normalbaselineskip+2pt]
& -21.75 & $<-6.40$\\[5pt]
& -20.25 & $<-6.10$\\[5pt]
& -19.00 & $<-5.00$\\[5pt]\hline

\end{tabular}
\end{table}

\subsection{Best-Fitting Parametric Functions}
\label{sec:fitting_results}

As described in Section \ref{sec:fitting}, we follow three different fitting approaches: Fitting a Schechter function, a DPL, and adopting the DPL parametrization from \citet{Donnan2024}, and only fitting for the normalization $\Phi^*$. We start with the Schechter fit at $z\sim10$, leaving all parameters unconstrained. In this case, of the 500 fits to the UVLF sampled from the \texttt{bagpipes} posterior, the majority prefers a solution with an extremely low normalization (log$(\Phi^*_{\rm Schechter})<-14$), and pushing ${\rm M^*_{UV,\,Schechter}}$ out to $<-30$, making the fitted curve look like a single power law with a slope of $\sim-3$. If we instead directly fit to the data points and their errors shown in Figure \ref{fig:uvlfs}, we find a decent fit with log$(\Phi^*_{\rm Schechter})=-6.47$, ${\rm M^*_{UV,\,Schechter}}=-22.75$, and $\alpha_{\rm Schechter}=-3.01$. The low normalization and high ${\rm M^*_{UV}}$ illustrate that our measured UVLF at $z\sim10$ is not well described by a Schechter function because it does not show an exponential cut-off at the magnitudes probed here. We therefore proceed with the DPL fit and find log$(\Phi^*_{\rm DPL})=-4.20_{-0.06}^{+0.23}$, ${\rm M^*_{UV,\,DPL}}=-20.12_{-0.18}^{+0.21}$, $\alpha_{\rm DPL}=-3.1_{-0.46}^{+0.96}$, and $\beta_{\rm DPL}=-3.36_{-0.49}^{+0.52}$. Both the bright and especially the faint end slope of the UVLF are poorly constrained in this fit due to the limited constraining power of the data.
Adopting the DPL parametrization proposed by \citet{Donnan2024} as ${\rm M^*_{UV,\,D+24}}(z=9.7)=-19.88$, $\alpha_{\rm D+24}(z=9.7)=-2.1$, and $\beta_{\rm D+24}(z=9.7)=-3.79$, we fit for the normalization and obtain log$(\Phi^*_{\rm D+24-fit})(z=9.7)=-3.79_{-0.19}^{+0.10}$, close to their value of log$(\Phi^*_{\rm D+24})=-3.72$. At $z\sim13$, the constraining power of our data is not sufficient to fit for all parameters of either the Schechter or the DPL curve simultaneously. We therefore only follow the third approach and find log$(\Phi^*_{\rm D+24-fit})(z=12.6)=-4.78_{-0.30}^{+0.62}$, significantly lower than the parametrization from  \citet{Donnan2024} which gives log$(\Phi^*_{\rm D+24})(z=12.6)=-4.12$.
Finally, at $z\sim17$, we perform the same fit only to our most constraining upper limit at M$_{\rm UV}\sim-20.25$ to get an upper limit on the normalization of log$(\Phi^*_{\rm D+24-fit})(z\sim17)<-5.09$, again lower than the \citet{Donnan2024} parametrization, log$(\Phi^*_{\rm D+24})(z=17)=-4.74$.

We show our fitted curves following the last approach, and the DPL fit at $z\sim10$ in Figure \ref{fig:uvlfs}, with the shaded regions highlighting the range between the 16th and 84th percentile of the fits to the 500 realizations of the UVLFs sampled from the \texttt{bagpipes} posterior.

\subsection{Comparing to Modeled UVLFs}
\label{sec:uvlf_modelling}

In the following, we compare our measurements to various theoretical models. Using the modeling framework from \citet{Shuntov2025b}, we first explore how changes to parameters tracing the star formation efficiency (SFE) and the burstiness of star formation can match our measured UVLFs. Then, we compare to models from the literature that implement changes to the IMF or the SFE at high redshift in a physically motivated way \citep{Hutter2025,Mauerhofer2025,Somerville2025}, and to a simple physical model without dust attenuation \citep{Ferrara2023}.

\subsubsection{The Shuntov et al. (2025) Framework}
\label{sec:shuntov_modelling}

In short, the \citet{Shuntov2025b} modeling framework is based on an instantaneous star formation efficiency (SFE) that relates the SFR to the halo mass accretion rate. The SFR is then converted to an M$_{\rm UV}$ assuming a conversion factor $\kappa_{\rm UV}$ from \citet{Madau2014} which is degenerate with $\epsilon_0$, the normalization of the SFE. The \citet{Shuntov2025b} framework assumes empirical relations for A$_{\rm V}-\beta$ \citep{Meurer1999} and $\beta-{\rm M_{UV}}$ \citep{Bouwens2014} to model dust, and adds Gaussian scatter to the $M_h-{\rm M_{UV}}$ relation, described by the parameter $\sigma_{\rm UV}$. They combine this with a model of the halo occupation distribution (HOD) to simultaneously model both the UVLF and the two-point correlation function (2PCF), and constrain the model parameters (i.e., the parametric form of $\epsilon_0$, $\sigma_{\rm UV}$ and the HOD) from UVLF and 2PCF measurements based on spectroscopic samples of [O\,{\sc iii}] and H$\alpha$ emitters from FRESCO \citep{Oesch2023} and CONGRESS at $z\sim4.3$, $z\sim5.4$, and $z\sim7.3$. Fixing the model parameters at the values constrained at $z=8$, they then extrapolate the UVLF to higher redshifts, essentially following the evolution of the halo mass function from \citet{Watson2013}.

In Figure \ref{fig:uvlfs}, we present their fiducial model with peak efficiency $\epsilon_0=0.23$ and scatter $\sigma_{\rm UV}=0.5$, as well as other models that illustrate the impact of different parameters on the UVLF. At $z\sim10$, the fiducial model only lies slightly below our measurements. Increasing the scatter to $\sigma_{\rm UV}=0.6$, and the efficiency to $\epsilon_0=0.35$ provides a decent fit to the data. To demonstrate the impact of dust on the UVLF, we show another model with $\sigma_{\rm UV}=0.5$ and $\epsilon_0=0.3$, but with no dust attenuation. This model overshoots at the bright end of the UVLF, emphasizing that in this framework, dust still plays an important role for the brightest galaxies at $z\sim10$. 

To match our measured UVLF at $z\sim13$, we further increase both parameters to $\epsilon_0=0.4$ and $\sigma_{\rm UV}=0.7$. Alternatively, a smaller peak efficiency $\epsilon_0=0.3$ and $\sigma_{\rm UV}=0.8$ with no dust attenuation produces a remarkably good fit to the data, illustrating the degeneracy between these different parameters in boosting the UVLF. While we cannot break these degeneracies, we emphasize that modest tweaks to the fiducial model are sufficient to reproduce our UVLF at $z\sim13$. 

In our highest redshift bin, at $z\sim17$, we show four models in addition to the fiducial one: (1) a ``maximum'' model that is just consistent with our upper limits, $\epsilon_0=0.7$, and $\sigma_{\rm UV}=1.4$, (2) the same without dust attenuation, which only has a minor impact at $z\sim17$, (3) one of the models shown to match the data at $z\sim13$ ($\epsilon_0=0.3$ and $\sigma_{\rm UV}=0.8$ without dust), and (4) a model with maximum efficiency and smaller UV-scatter ($\epsilon_0=1.0$, $\sigma_{\rm UV}=0.5$, no dust). The latter two lie well below our upper limits. A further increase in $\epsilon_0$ and/or $\sigma_{\rm UV}$ with redshift at $\gtrsim13$ is thus allowed by the data, but not required.

\subsubsection{Other Models from the Literature}
\label{sec:lit_models}

Motivated by local observations of a more top-heavy IMF in regions with higher SFR \citep[e.g.][]{Jerabkova2018}, and the idea that denser gas regions are more likely to form massive stars, \citet{Hutter2025} model an IMF that becomes more top-heavy with increasing specific SFR (sSFR) above a redshift-dependent threshold in sSFR. The curves shown in Figure \ref{fig:uvlfs} represent two models: one with a Salpeter IMF, and one with their evolving IMF. Both models lie slightly above our measurements at $z\sim10$, where they produce very similar UVLFs. At $z\sim13$, their two models differ more significantly. Interestingly, the model based on a Salpeter IMF is consistent with the data, while the evolving IMF model lies above our measurement at ${\rm M_{UV}}\sim-20$. Both models are consistent with our upper limits at $z\sim17$, with the evolving IMF model being close to our 2$\sigma$ upper limits at ${\rm M_{UV}}\sim-19$ and ${\rm M_{UV}}\sim-20$. The relatively rapid evolution in the UVLF from $z\sim10-17$ we see in the data is therefore more consistent with their Salpeter IMF model, than their evolving IMF scenario.

We further show models from \citet{Mauerhofer2025} who implement interstellar medium physics from \texttt{SPHINX$^{20}$} \citep{Rosdahl2022} into the \texttt{DELPHI} semi-analytical model \citep{Dayal2014}. We show their fiducial model, one where they implement an increasing SFE with redshift (eSFE), and one with an evolving IMF (eIMF) that becomes more top-heavy at lower metallicity and higher redshift. Their fiducial model lies below our measured UVLFs at the bright end at $z\sim10$, and $z\sim13$, meaning that a boost to these modeled UVLFs is required to match the data. Indeed, both the eSFE and the eIMF models provide better fits at $z\sim10$, with the eIMF being slightly too high at the faint end. The eIMF model agrees with our data points at $z\sim13$ within error bars, while the eSFE model lies above our constraints at M$_{\rm UV}\sim-20$. At $z\sim17$, all their models are consistent with our constraints, with both the eIMF, and the eSFE models being close to our upper limit at ${\rm M_{UV}}\sim-19$ due to their steepness. None of their models therefore consistently reproduces our measured UVLFs across the redshift and magnitude range explored here. While their eSFE and eIMF models match the measured number density in the brightest bins of the UVLF at $z\sim10-13$, they tend to produce too many galaxies at fainter magnitudes, and possibly at $z\gtrsim15$.

Next, we compare to models from \citet{Somerville2025} who implement a scaling of the SFE with the gas surface density into the Santa Cruz semi-analytical model \citep{Somerville1999,Somerville2001}. This model is labeled as density-modulated star formation efficiency (DMSFE). We show their baseline model, two versions of the DMSFE model with different fractions of the ISM in dense star forming clouds ($f_d=0.1$, and $f_d=0.5$), and the DSFME model with $f_d=0.1$ and additional bursts in star formation added in post-processing. At $z\sim10$, their baseline model is consistent with our measurements down to M$_{\rm UV}\sim-21$, but produces higher number densities in the brightest bin (M$_{\rm UV}\sim-22$). All other models produce even higher UVLFs. Note that the models shown here are not corrected for dust attenuation, whose impact is however limited (see e.g., Figure 10 in \citealt{Somerville2025}). At $z\sim13$, their baseline model lies slightly below the data at M$_{\rm UV}\sim-20$, while the DMSFE with $f_d=0.1$ provides a good fit across the entire magnitude range. The other two modeled UVLFs ($f_d=0.5$, and $f_d=0.1$ with additional bursty star formation) still lie above our measurements. The baseline and the $f_d=0.1$ models lie below our upper limits at $z\sim17$, where the $f_d=0.5$ model lies slightly above our upper limit at M$_{\rm UV}\sim-19$. Consistent with the models explored previously, this indicates that only a modest change to the fiducial models (in this case, an ISM fraction in dense clouds of 10\%) is needed to account for the number density of UV-bright galaxies at $z\sim13$, and substantially more significant changes to the baseline model are disfavored by both our measured UVLF at $z\sim13$ and the upper limits at $z\sim17$.

Finally, we compare to the attenuation free model (AFM) from \citet{Ferrara2023} who explore a minimal physical model where high redshift galaxies are not attenuated by dust. This could be due to efficient feedback-driven ejection of dust in early galaxies. At $z\sim10$, their model produces slightly higher number densities at the bright end compared to our measurements, but it provides a good fit at $z\sim13$, and lies well below our upper limits at $z\sim17$. In the context of this simple model, the slightly lower number densities we measure at the bright end at $z\sim10$ may be related to non-negligible dust attenuation affecting the brightest galaxies at those epochs, as also suggested by our modeling based on the framework from \citet{Shuntov2025b}.

\subsection{UV Luminosity and Star Formation Rate Density}
\label{sec:density}

\begin{figure*}
    \centering
        \includegraphics[width=0.9\textwidth]{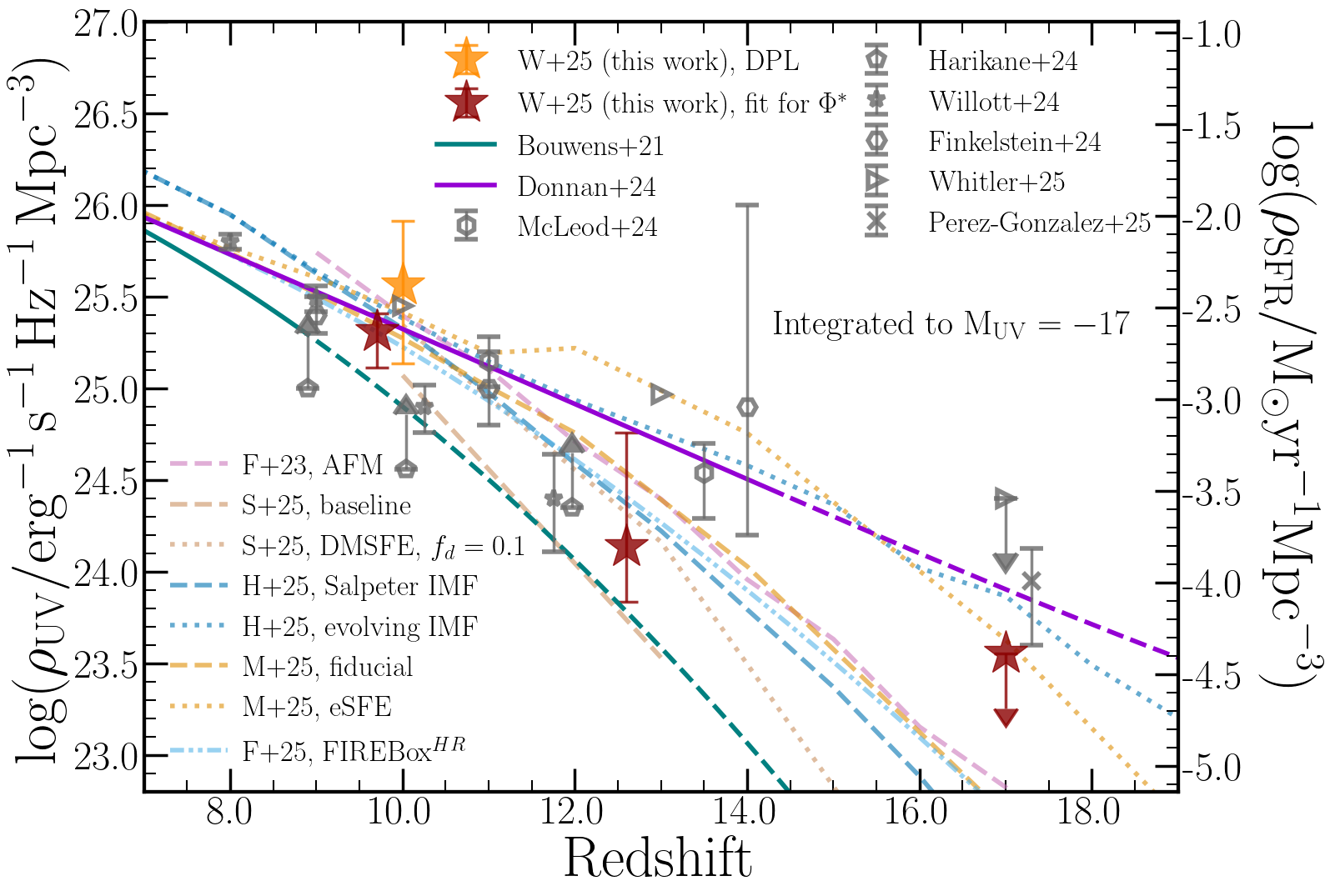}
    \caption{UV luminosity density obtained from integrating the fitted parametric curves, compared to various results from the literature \citep{McLeod2024,Harikane2024,Willott2024,Finkelstein2024,Whitler2025, PerezGonzalez2025}, and theoretical work, where we show a subset of the models displayed on the right of Figure \ref{fig:uvlfs}: the AFM from \citet{Ferrara2023} (F+23), the baseline and the DMSFE model with $f_d=0.1$ from \citet{Somerville2025} (S+25), the Salpeter and the evolving IMF model from \citet{Hutter2025} (H+25), as well as the fiducial and the eSFE model from \citet{Mauerhofer2025} (M+25). In addition, we show the theoretical model based on the SFE-$M_{\rm halo}$ relation of the FIREBox$^{HR}$ simulation from \citet{Feldmann2025} (F+25)}. The parametrizations from \citet{Bouwens2021} and \citet{Donnan2024} are shown as the teal and dark violet line respectively, where the extrapolated part is dashed. The secondary y-axis shows the cSFRD, based on the conversion factor from \citet{Madau2014}.
    \label{fig:rho_UV}
\end{figure*}

We integrate our fitted parametric curves in the range M$_{\rm UV}\in(-23,-17)$ to calculate the cosmic UV-luminosity density $\rho_{\rm UV}$, and show it in Figure \ref{fig:rho_UV}.
To estimate uncertainties, we integrate the curves fitted to the 500 realizations of our UVLFs, and show the median, 16th and 84th percentile of the resulting 500 values of $\rho_{\rm UV}$ as our measurement and error bar respectively. Our $\rho_{\rm UV}$ measurements and uncertainties are listed in Table \ref{tab:rho_uv_values}.

Adopting the conversion factor from \citet{Madau2014}, $\kappa_{\rm UV}=1.15\times10^{-28}{\rm M_\odot\,yr^{-1}/(erg\,s^{-1}\,Hz^{-1})}$, we show the cosmic star formation rate density (cSFRD) on the secondary y-axis on the right of Figure \ref{fig:rho_UV}. We compare to JWST-based literature results from \citet{McLeod2024,Harikane2024,Willott2024,Finkelstein2024,Whitler2025}, and \citet{PerezGonzalez2025}, as well as to a subset of the models displayed on the right of Figure \ref{fig:uvlfs}, where we consistently integrate the modeled UVLFs in the range M$_{\rm UV}\in(-23,-17)$ to enable a direct comparison to the data. We further show a pre-JWST parametrization and its extrapolation to $z>9$ from \citet{Bouwens2021}, the parametrization from \citet{Donnan2024}, and a theoretical model from \citet{Feldmann2025} that is based on the weakly mass-dependent SFE-$M_{\rm halo}$ relation found in the FIREBox$^{HR}$ simulation.

\begin{table}
\centering
\caption{Measured values of the UV luminosity density in our three redshift bins, obtained by integrating our best-fitting DPL curves (adopting the parametrization from \citet{Donnan2024}, and fitting for $\Phi^*$) down to M$_{\rm UV}=-17$.}
\label{tab:rho_uv_values}
\begin{tabular}{ccc}
\hline
& & \\[\dimexpr-\normalbaselineskip+1pt]
median redshift & log($\rho_{\rm UV}/{\rm erg^{-1}\,s^{-1}\,Hz^{-1}\,Mpc^{-3}})$ \\
& & \\[\dimexpr-\normalbaselineskip+1pt]
\hline
\hline
& & \\[\dimexpr-\normalbaselineskip+2pt]
9.7 & $25.31^{+0.10}_{-0.19}$\\[5pt]
12.6 & $24.14^{+0.62}_{-0.30}$\\[5pt]
17 & $<23.55$\\[5pt]

\hline

\end{tabular}
\end{table}

As expected, at $z\sim10$, our measurements are consistent with the displayed JWST-based literature results. The values and error bars inferred from the DPL fit (orange star) illustrate how the limited constraining power of our data at the faint end of the UVLF increases the uncertainty in $\rho_{\rm UV}$. While these uncertainties are in part driven by fits that match the bright end of the UVLF, but are inconsistent with constraints from the literature at the faint end (see the orange shaded regions in Figure \ref{fig:uvlfs}, top left panel), fixing the DPL slopes and turnover magnitude to the values found by \citet{Donnan2024} likely causes us to underestimate the true uncertainty in $\rho_{\rm UV}$. Nevertheless, our measurement at $z\sim13$, and the upper limit at $z\sim17$ indicate a somewhat more rapid evolution of the cSFRD at $z\sim10-17$ than found previously, while still being consistent with literature values within error bars at $z\sim13$. Our 2$\sigma$ upper limit at $z\sim17$ is just below the $1\sigma$ uncertainty from \citet{PerezGonzalez2025}, indicating a more rapid drop of the cSFRD beyond $z\sim14$ than suggested by their measurement. 

All models presented in Figure \ref{fig:rho_UV} are consistent with our measurement at $z\sim10$, in particular if the generous error bars from the full DPL fit are considered. We note that we only show a subset of the models shown in Figure \ref{fig:uvlfs}: the fiducial models, and those that best match our UVLFs from \citet{Somerville2025}, \citet{Mauerhofer2025} and \citet{Hutter2025} respectively. At $z\sim13$, most models are consistent with our measurement, with the baseline model from \citet{Somerville2025} lying below and the eSFE model from \citet{Mauerhofer2025} above. The latter, as well as the evolving IMF model from \citet{Hutter2025} lie above our upper limit at $z\sim17$. Models that consistently reproduce our measured UV luminosity density across all redshifts are the AFM from \citet{Ferrara2023}, the DMSFE model with $f_d=0.1$ from \citet{Somerville2025}, the Salpeter IMF model from \citet{Hutter2025}, the fiducial model from \citet{Mauerhofer2025}, and the model from \citet{Feldmann2025}. In line with our discussion of the modeled UVLFs above, only small tweaks to the fiducial models, if at all, are required to match the abundance of galaxies we observe at $z\gtrsim10$. 

\section{Discussion}
\label{sec:discussion}

In summary, we find UVLFs that are consistent with JWST-based literature results at $z\sim10$, and confirm the high abundance of UV-bright (M$_{\rm UV}\sim-22$) galaxies with better number statistics and mitigated cosmic variance effects. We then find somewhat lower number densities at $z\sim13$, especially at M$_{\rm UV}\sim-20$, and provide the most constraining upper limits on the number density at $z\sim17$ to date. Together, these measurements suggest a more rapid evolution in the abundance of galaxies, and the UV luminosity density, than found previously. Comparing these findings to theoretical models shows that fiducial models that match the UVLF at $z\sim10$, only need to be tweaked slightly to match our constraints at $z\sim13$ and beyond. We discuss some remaining caveats and limitations of this study below, before providing some additional interpretation of our results.

\subsection{Caveats and Limitations}

\subsubsection{Inhomogeneous Filter Coverage}
\label{sec:inhomogeneous_coverage}

While our color-based sample selection described in Section \ref{sec:sample_selection} largely relies on the NIRCam wide filters F115W, F150W, F200W, F277W, F356W, and F444W that are available across all fields used in this work, the inhomogeneous coverage by other filters nevertheless affects our results. 
First, the F410M filter is available across all legacy fields, as well as for the deepest five PANORAMIC pointings. Further, additional medium band filters are available for parts of the GOODS fields as well as the EGS, UDS, and COSMOS fields and the A2744 cluster (see Section \ref{sec:imaging} for a list of programs that contributed imaging to these fields).
The additional filters help to break degeneracies in the \texttt{eazy} fits, and therefore enable a more robust removal of low-z interlopers based on the $z_{\rm phot}$ probability distribution function.

Second, filters probing wavelengths below the Lyman break at $z\gtrsim9$ are available from JWST (F090W for all the legacy fields, F070W for a small fraction), and HST (ACS and WFC3 across most of the legacy fields). They help rule out low-z interlopers in two ways: again, by improving the photometric redshift constraints, and additionally during the visual inspection, by identifying sources with clearly visible flux in filters below the supposed Lyman break.
The availability of additional filters therefore improves the quality of our selection over certain parts of the fields, in particular the legacy fields. At the same time, we would expect a higher interloper fraction in PANORAMIC pointings with imaging in only six wide filters. 

As a test on the robustness of ``6-filter-only'' photometric redshifts at $z\sim10$, we run \texttt{eazy} on the part of the Abell-2744 cluster field that is covered by the Mega Science program, meaning that all NIRCam medium bands, as well as HST/ACS filters are available. We take the resulting photometric redshifts that have been shown to be robust out to at least $z\sim7$ in \citet{Naidu2024} through NIRCam/grism spectroscopy as the ground truth, and then re-run \texttt{eazy} on just the PRIMER filter set (HST/ACS + 7 NIRCam wide filters from F090W to F444W, and F410M), and on just the PANORAMIC filter set (6 NIRCam wide filters from F115W - F444W). Applying our F115W dropout selection, we identify 9 high-z candidates, 3 of which are confidently fit at low redshift with both the full and the PRIMER filter set. 2 out of those 3 with low-z solutions are fit at $z\gtrsim9$ when using just the PANORAMIC filter set. This means that if we had selected F115W dropouts using the PANORAMIC filter set, we would have selected 8, 2 of which are low-z interlopers, corresponding to a contamination fraction of 25\%. For comparison, the HST pure parallel program BoRG \citep{Bradley2012} specified contamination fractions ranging from $\sim40$\%\ to $\sim90$\%\ for dropout-selected $z\sim9-10$ galaxies at H-band magnitudes of 24.8 to 26.5, highlighting the substantial improvement that JWST provides in selecting $z\sim9-10$ galaxies from pure parallel imaging. 

We note that possible remaining interlopers in our sample would boost the measured UVLF. Therefore, removing any interlopers would further strengthen our conclusion of relatively low number densities compared to other recent literature results, in particular at $z\sim13$.

\subsubsection{$z\sim13$ Candidates with Extended Morphology}
\label{sec:ext_z13_candidates}

\begin{figure}
    \begin{center}
        \includegraphics[width=0.87\columnwidth]{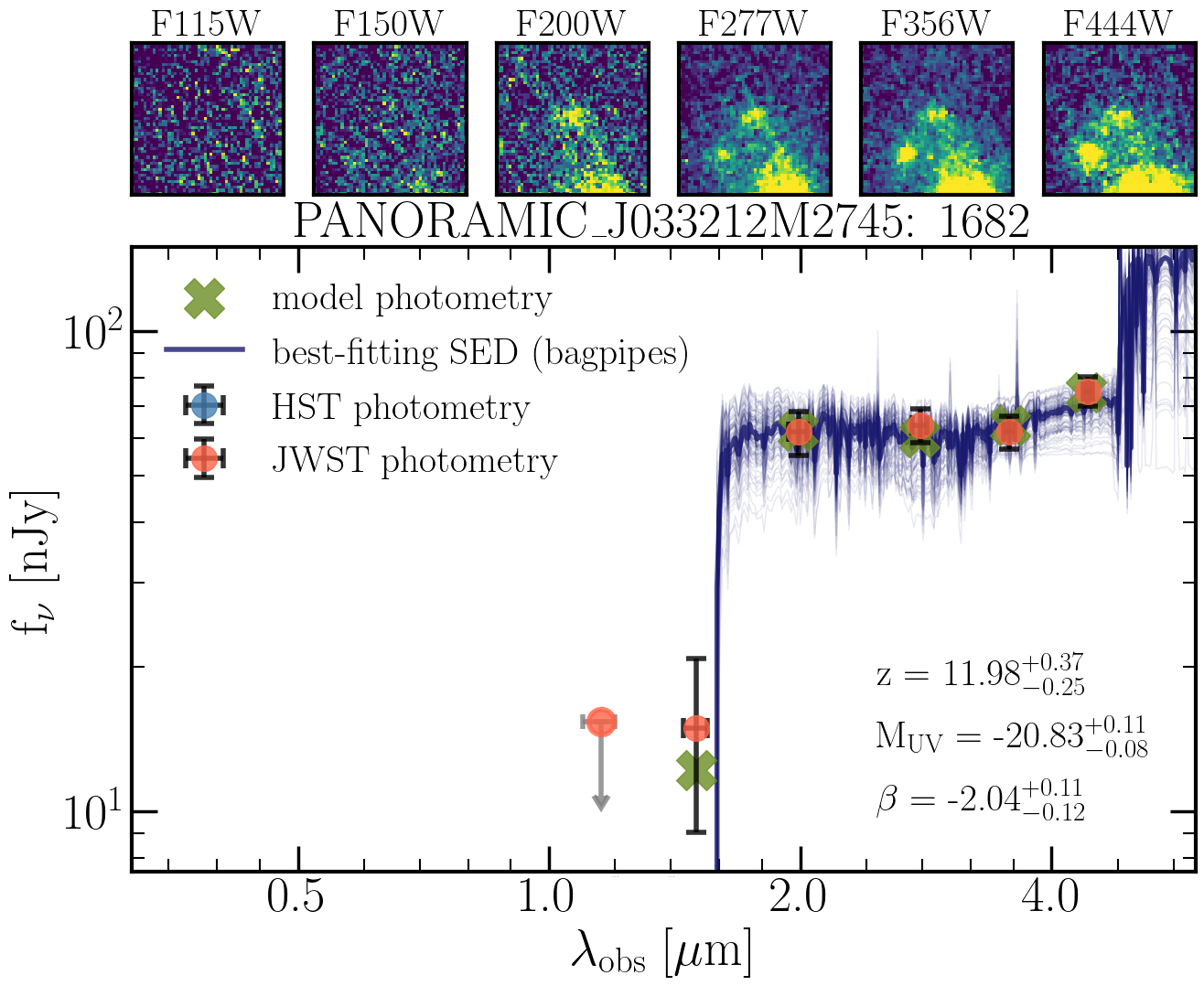}
        \newline
        \includegraphics[width=0.87\columnwidth]{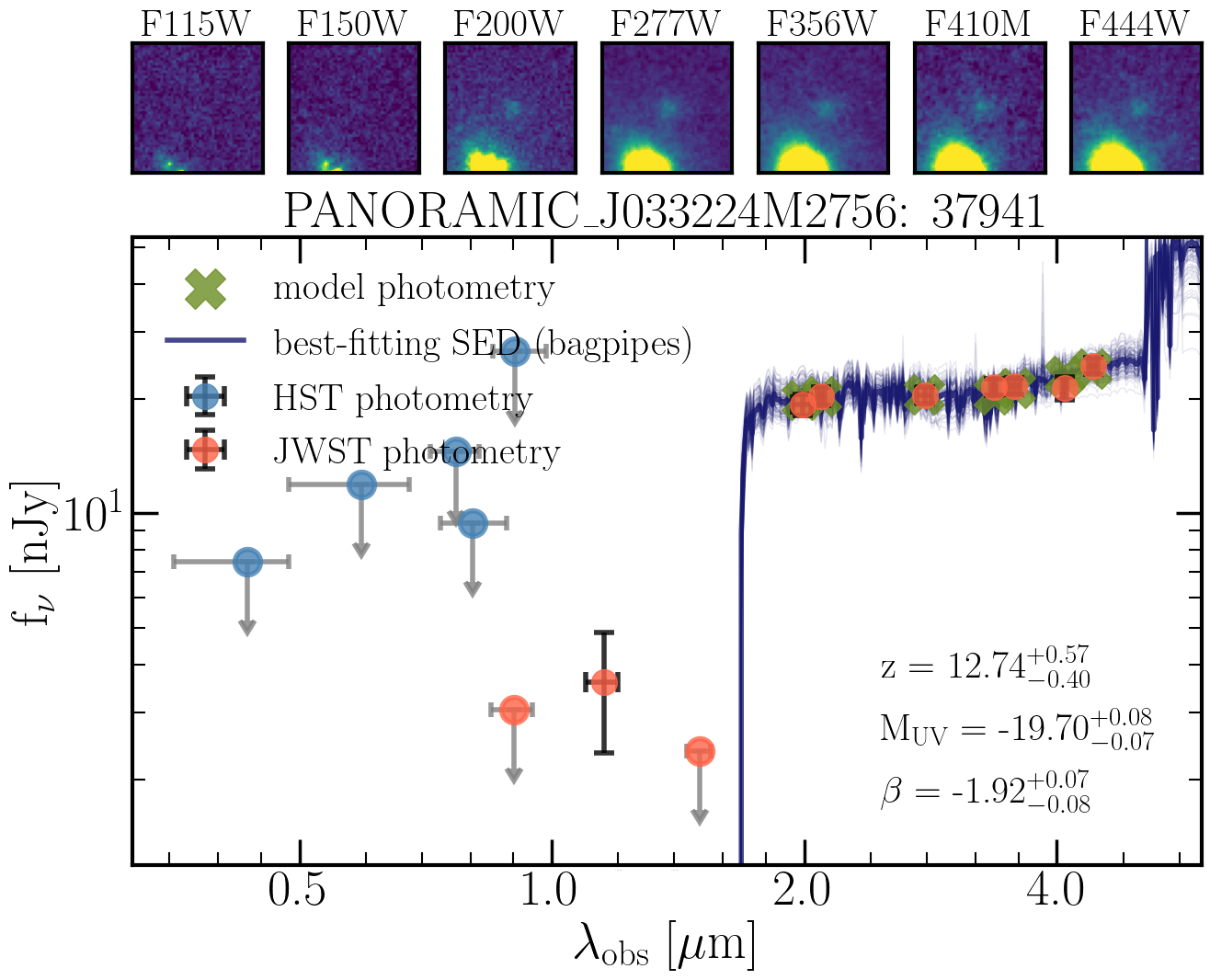}
        \newline
        \includegraphics[width=0.87\columnwidth]{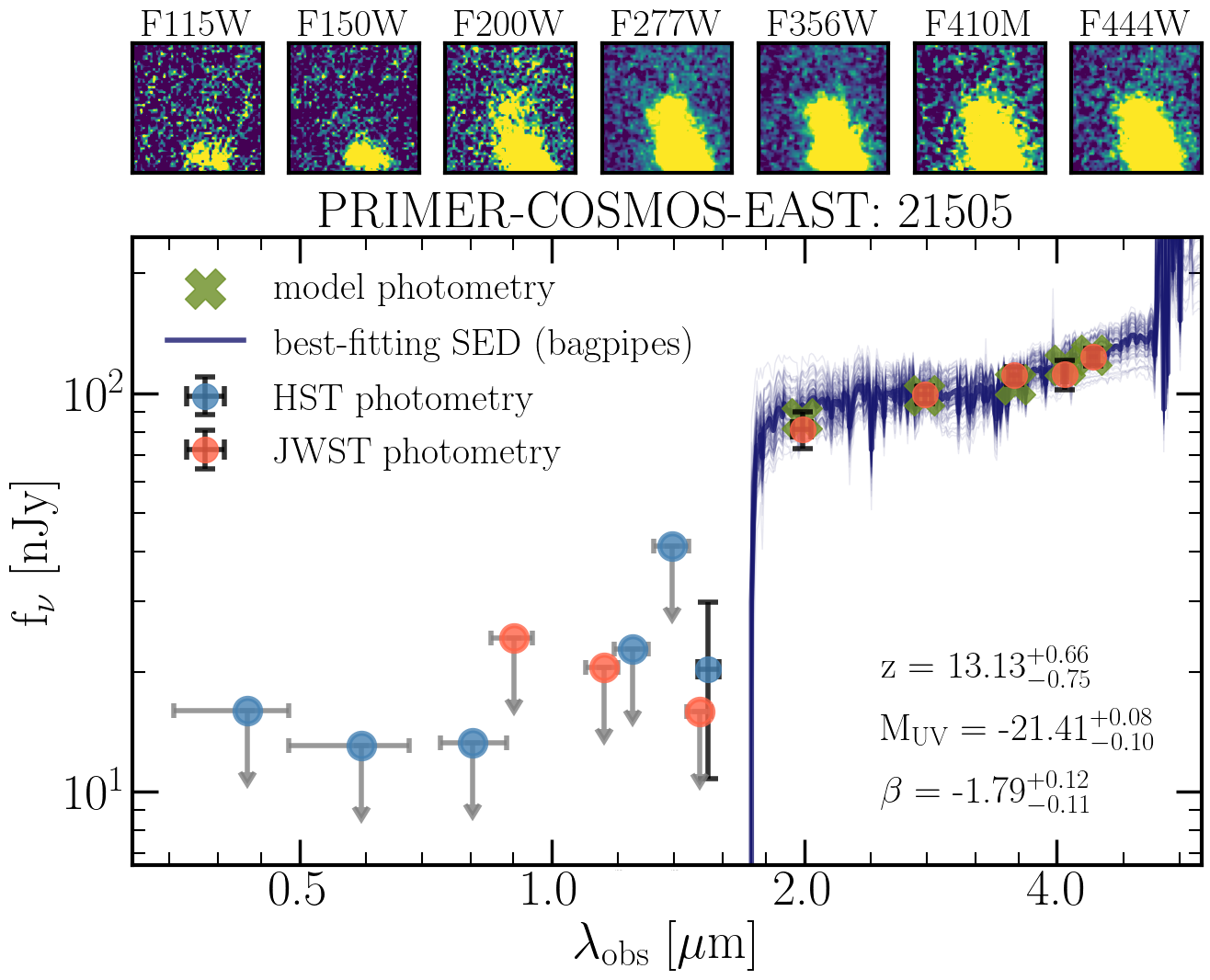}
    \end{center}
    \caption{Three F150W dropout $z\sim13$ candidates from PANORAMIC and PRIMER-COSMOS, in analogy to Figures \ref{fig:z10_candidates} and \ref{fig:z13_candidates}. They all show extended morphologies, and relatively low surface brightness, and they are located next to a bright galaxy whose $z_{\rm phot}$ is consistent with $z\sim3.4-3.5$, a typical contaminant redshift of our F150W dropout sample.}
    \label{fig:fuzzy_z13_galaxies}
\end{figure}

During visual inspection, we identified 3 galaxies in our $z\sim13$ sample that are confidently fit at high redshift, and show an extended morphology and relatively low surface brightness. We show the photometry, SEDs and imaging cutouts of the three sources in Figure \ref{fig:fuzzy_z13_galaxies} in analogy to Figures \ref{fig:z10_candidates} and \ref{fig:z13_candidates}.

All of them are located close to a brighter galaxy whose photometric redshift from \texttt{eazy} is consistent with $z\sim3.4-3.5$. This redshift is common among low-redshift interlopers in our $z\sim13$ sample, due to H$\beta$ and [O\,{\sc iii}] boosting F200W, and H$\alpha$ boosting F277W, and F356W, while F444W roughly probes the rest-frame optical to near-infrared continuum, mimicking a blue UV-continuum at $z\sim13$. 

We remove ID 21505 from our sample before computing the UVLF, as it is the brightest, and most extended of the three, while also being most strongly blended with the brighter neighboring galaxy. Further, it shows moderately red colors in the supposed UV-continuum, with ${\rm F200W}-{\rm F444W}=0.46$ which is close to the edge of our selection box at 0.5. This results in a UV-slope of $\beta=1.76\pm0.12$ which is atypical for galaxies at $z\sim13$ \citep[e.g.][]{Topping2024}. Including this object would further boost the number density in the brightest bin at $z\sim13$.
We keep the other two sources in our sample and note here that removing them would yield a lower number density in the intermediate bin at $z\sim13$ (M$_{\rm UV}\sim-20$).

\subsubsection{Sparse Sampling of the Selection Function}
\label{sec:selection_function}

In our completeness calculation, we assign completeness values to each source by sampling intrinsic properties (M$_{\rm UV}$, $z$, and $\beta$) from the \texttt{bagpipes} posterior distribution, and then simulating the selection by creating mock photometry and adding noise to each filter according to the noise in our photometric catalog. The resulting completeness depends on the intrinsic properties, and the noise properties of the images in different filters. This statistically provides accurate correction factors for the effective volume from which we select galaxies. For example, when computing our UVLFs, we divide by the volume given by the survey area in a given bin of M$_{\rm UV}$ (Section \ref{sec:survey_area}), and the nominal redshift range of our sample (Section \ref{sec:sample_selection}). In practice, the effectively sampled volume is smaller due to the incompleteness of our selection towards the edges of the nominal redshift range.

Accurately correcting for the incompleteness requires sufficient number statistics to sample the selection function along several dimensions (M$_{\rm UV}$, $z$, $\beta$, and in principle, the noise level in different filters that are part of our selection). A single ``outlier'' at the edge of the nominal redshift range can be assigned a low completeness and significantly boost the UVLF. Conversely, the lack of such galaxies can lead to an overestimation of the effective survey volume and thus an underestimation of the UVLF. The same is true for the noise properties. Galaxies selected close to the selection threshold get assigned low values of completeness which are statistically correct, but can significantly affect the UVLF for small sample sizes. We mitigate effects of low completeness values by capping the inferred selection completeness at a somewhat arbitrary value of 0.02. Changing this value only significantly affects the inferred number density in the faintest M$_{\rm UV}$ bin respectively. Lowering it below 0.02 slightly boosts the number density in that bin, while also increasing the uncertainty. We argue that completeness values $\lesssim0.02$ cannot be reliably inferred given our setup, and that this problem can only be solved with larger samples of faint galaxies at high redshift, which are not the focus of this work. The number densities we measure at M$_{\rm UV}\lesssim-20$ are largely insensitive to the completeness correction.

\subsection{A Rapidly Evolving Galaxy Population at $z>10$?}

Compared to previous literature results, we find that the number density of galaxies at M$_{\rm UV}\lesssim-18.5$ drops relatively rapidly from $z\sim10$ to $z\sim13$ and beyond. This is also reflected in a more rapid decline of the UV luminosity density (Figure \ref{fig:rho_UV}) where our results are however still consistent with most literature results within uncertainties. This is mostly because $\rho_{\rm UV}$ depends on the faint-end slope of the UVLF which is not well constrained by our data, and sensitive to the completeness correction discussed in the previous Section. For our fiducial measurements, we have adopted the UVLF parametrization from \citet{Donnan2024} which gives $\alpha_{\rm DPL}=-2.1$ at $z=12.6$. We have tested that if we instead assume a steeper faint-end slope of $\alpha_{\rm DPL}=-2.6$, as found e.g., at $z\sim10$ in \citet{Whitler2025}, this only increases our inferred UV-luminosity density by $\sim0.1$ dex at $z\sim13$ which does not significantly change our finding of a more rapid decline compared to other results.

We note however, that in the brightest bin at $z\sim13$, we measure a number density that is consistent with e.g., \citet{Donnan2024} who measured it at $z\sim12.5$. Moreover, 3 out of 4 galaxies in that bin lie at photometric redshifts $>14$, with one of them spectroscopically confirmed at $z=14.1796$ \citep{Carniani2024,Carniani2024b}, and another one at $z=13.53$ \citep{Donnan2026}. This indicates that the number density we would infer at the bright end of the UVLF would be similar at $z\sim14$, suggesting a slow evolution in the number density of the very brightest galaxies (M$_{\rm UV}\lesssim-21$). This further results in a shallow bright-end slope of the UVLF which is also seen at $z\sim10$ where a Schechter function does not provide a good fit to the data due to the lack of an exponential cut-off out to M$_{\rm UV}\sim-22$. Spectroscopic follow-up of the remaining bright candidates at $z\sim14$, and deep JWST imaging data over larger areas, which may be obtained efficiently through pure parallel imaging in future cycles of JWST observations, are required to shed more light on the rapid build-up of these surprisingly luminous sources.

Our UVLFs are then qualitatively consistent with the picture outlined in \citet{Donnan2025}, where the abundance of galaxies at the bright end of the UVLF is related to the ever younger ages of the stellar populations in galaxies at $z\gtrsim10$. In terms of the modeling of the UVLF from \citet{Shuntov2025b} (Section \ref{sec:shuntov_modelling}), it is the parameter $\kappa_{\rm UV}$, used to convert between the SFR and UV luminosity, that is usually assumed to be fixed to the value from \citet{Madau2014}, which is based on a stellar population age of $100\,$Myr. As we approach the first galaxies, the average age of their stellar populations is however expected to drop below $100\,$Myr, which then boosts the UV luminosity at a given SFR.

The average ages assumed in \citet{Donnan2025} suggest a formation redshift of $z\sim15$, meaning that they predict a rapid decline in the number density of galaxies at $z>15$, following the rapid evolution of the halo mass function. This may be in line with the relatively rapid decline we find from $z\sim10$ to $z\sim13$ and certainly with the lack of candidates at $z\sim17$. Whether this scenario can also account for the presence of the brightest galaxies at $z\sim14$ remains to be investigated.

\section{Summary and Conclusions}
\label{sec:summary_conclusions}

Combining NIRCam imaging data from legacy fields and the pure parallel program PANORAMIC, we perform measurements of the UVLF at $z\gtrsim9$ over the largest survey area to date ($0.28\,{\rm deg}^2$) with deep NIRCam imaging in 6 or more filters, and along 35 independent lines of sight. 

We start by selecting robust photometric candidates using a color-color selection that requires a strong Lyman break ($>1.5\,$mag with a 2$\sigma$ upper limit in the dropout filter), and a ${\rm SNR}>8$ redward of the break. After removing confident low redshift interlopers based on photometric redshifts from \texttt{eazy}, and visually vetting our samples, we identify 86 F115W dropouts ($z\sim10$), 10 F150W dropouts ($z\sim13$) and no robust F200W dropouts ($z\sim17$). 

We perform SED-fitting with \texttt{bagpipes} to infer photometric redshifts and UV properties for our entire galaxy sample. Sampling from the resulting posterior distributions, taking into account the incompleteness of our samples, as well as measuring the survey area as a function of the M$_{\rm UV}$ limit that can be probed across all survey fields, we infer UVLFs, as well as the UV-luminosity density at $z\sim10$, $z\sim13$, and $z\sim17$. 

Our findings can be summarized as follows,

\begin{itemize}
\item Cross-matching with publicly available NIRSpec spectra, we find that 36 of our F115W and F150W dropout candidates have robust spectroscopic redshifts, and all but one of them are unambiguously confirmed to be at $z\gtrsim9$. Two of our F200W dropout candidates are spectroscopically confirmed to be low-redshift interlopers, one of them a previously unpublished source at $z_{\rm spec}=4.9$ from CAPERS.
\item We identify 14 new and robust candidates at $z\sim10$, as well as two candidates at $z\sim14$ from PANORAMIC pure parallel imaging, one of which is now confirmed to be at $z=13.53$ \citep{Donnan2026}. If confirmed, the remaining $z\sim14$ candidate from PANORAMIC may be the brightest source known at this redshift.
\item Our UVLF at $z\sim10$ is consistent with a wealth of literature results, confirming a high abundance of UV-bright galaxies (M$_{\rm UV}\lesssim-21$) with better number statistics. At $z\sim13$, we find somewhat lower number densities compared to most literature results, in particular at M$_{\rm UV}\sim-20$. This suggests a rapid evolution in the number density from $z\sim10-13$, reflected in a drop by an order of magnitude in the UV luminosity density over these redshifts.
\item Our upper limits on the UV luminosity density at $z\sim17$ are the most constraining limits available to date, yielding log($\rho_{\rm UV}/{\rm erg}^{-1}{\rm s}^{-1}{\rm Hz}^{-1}{\rm Mpc}^{-3})<23.60$, which implies that $\rho_{\rm UV}$ increases by at least a factor $\sim50$ from $z\sim17$ to $z\sim10$.
\item This indicates a rapid build-up of the galaxy population from $z\gtrsim15$ to the surprising presence of M$_{\rm UV}\sim-21$ galaxies at $z\sim14$, and the high abundance of even more luminous galaxies at $z\sim10$. 
\item Comparing these findings to models and simulations suggests that only a modest boost to fiducial modeled UVLFs in the recent literature is required to match the observed number densities of galaxies at $z\gtrsim10$. This boost is achievable through enhanced burstiness of star formation, a more top-heavy IMF, a higher star formation efficiency, a lack of dust attenuation, or a combination thereof, perhaps limited to galaxies or regions within galaxies with high gas densities.
\end{itemize}

We have illustrated the power of combining deep legacy imaging data over large contiguous fields with pure parallel imaging that adds many independent lines of sight to constrain the properties of the galaxy population at $z\gtrsim9$. The large available area allows us to apply conservative color-based selection cuts, yielding robust samples of high redshift galaxies while maintaining reasonable number statistics. The independent lines of sight not only mitigate the effect of cosmic variance on the measured number density of galaxies, but also allow for a direct measurement of the clustering strength at $z\sim10$, which can help discriminate between different models proposed to explain the abundance of UV-bright galaxies at $z\gtrsim10$. We explore this in a companion paper \citep{Weibel2025}. 

The measurements presented here are based on the largest currently available imaging area that enables the robust high-z selections we performed. Further breakthroughs on the bright end of the UVLF at $z\gtrsim9$ will require larger samples, that must be collected from ever-larger deep NIRCam imaging area. In the future, such expansion of area will be a critical resource for confirming our finding of a rapid decline in the abundance of galaxies at $z>10$, and in placing stronger constraints at $z\gtrsim14$, where only very few, but remarkably bright galaxies are known to date. At the same time, identified candidates must be followed up spectroscopically, to test and confirm photometry-based results and to better characterize the still poorly understood populations of low redshift interlopers contaminating dropout searches at $z\gtrsim11$.

\section*{Acknowledgements}

This work is based on observations made with the NASA/ESA/CSA James Webb Space Telescope. The data were obtained from the Mikulski Archive for Space Telescopes at the Space Telescope Science Institute, which is operated by the Association of Universities for Research in Astronomy, Inc., under NASA contract NAS 5-03127 for JWST. These observations are associated with program \#2514.
The Cosmic Dawn Center is funded by the Danish National Research Foundation (DNRF140). This work has received funding from the Swiss State Secretariat for Education, Research and Innovation (SERI) under contract number MB22.00072, as well as from the Swiss National Science Foundation (SNSF) through project grant 200020\_207349. The work of CCW is supported by NOIRLab, which is managed by the Association of Universities for Research in Astronomy (AURA) under a cooperative agreement with the National Science Foundation.  Support for program JWST-GO-2514 was provided by NASA through a grant from the Space Telescope Science Institute, which is operated by the Association of Universities for Research in Astronomy, Inc., under NASA contract NAS 5-03127. 

\facilities{JWST(NIRSpec, NIRCam)}

\software{
All software packages used in this work  are publicly available on Github: \texttt{grizli}, \texttt{msafit}, \texttt{msaexp}. We acknowledge: 
    astropy \citep{2013A&A...558A..33A,2018AJ....156..123A,2022ApJ...935..167A}, 
    matplotlib \citep{10.1109/MCSE.2007.55},  
    numpy \citep{10.1038/s41586-020-2649-2}, 
    scipy \citep{10.1038/s41592-019-0686-2}, 
    \texttt{cosmic-variance} \citep{Jespersen2025},
    the \texttt{jwst} pipeline (\citealt{10.5281/zenodo.10870758}), 
    \texttt{msaexp} (\citealt{10.5281/zenodo.7299500}), 
    and \texttt{grizli} (\citealt{10.5281/zenodo.1146904})}

\bibliography{paper}{}

\bibliographystyle{aasjournal}

\appendix

\section{Tables of High Redshift Candidates}
\label{sec:candidate_tables}

We list our F115W dropout ($z\sim10$) and F150W dropout ($z\sim13$) candidates in tables \ref{tab:z10_candidates} and \ref{tab:z13_candidates}. For each candidate we specify the ID, field, position, UV-magnitude (corrected for lensing magnification in the A2744 field), UV-slope, photometric redshift from \texttt{bagpipes}, the spectroscopic redshift from the DJA where available, and the corresponding reference (paper or program ID).
\clearpage

\LTcapwidth=\textwidth
\begin{longtable*}{ccccccccc}
\caption{$z\sim10$ F115W dropout candidates with ID, field, R.A, Declination, M$_{\rm UV}$, UV-slope $\beta$, z$_{\rm phot,\,bp}$ from \texttt{bagpipes} where the redshift is constrained to the nominal range of the dropout selection, z$_{\rm spec}$ from the DJA where available, and a corresponding reference or JWST program ID. The UV-magnitudes in the A2744 cluster field are corrected for lensing magnification (see Section \ref{sec:lensing}).}
\label{tab:z10_candidates}
\\& & & & & & & &\\[\dimexpr-\normalbaselineskip+0pt]
\hline
& & & & & & & &\\[\dimexpr-\normalbaselineskip+3pt]
ID & Field & R.A. & Decl. & M$_{\rm UV}$ & UV-slope $\beta$ & \shortstack{z$_{\rm phot,\,bp}$} & z$_{\rm spec}$ (DJA) & z$_{\rm spec}$ Reference\\
& & & & & & & &\\[\dimexpr-\normalbaselineskip+3pt]
\hline
\hline
4811 & A2744 & 3.59303 & -30.44976 & $-19.29^{+0.05}_{-0.05}$ & $-2.66^{+0.04}_{-0.06}$ & $10.63^{+0.16}_{-0.20}$ & - & -\\[5pt]
4995 & A2744 & 3.63757 & -30.44915 & $-20.76^{+0.07}_{-0.11}$ & $-2.61^{+0.06}_{-0.05}$ & $10.15^{+0.27}_{-0.74}$ & - & -\\[5pt]
10719 & A2744 & 3.56537 & -30.43232 & $-19.36^{+0.13}_{-0.09}$ & $-2.65^{+0.04}_{-0.05}$ & $9.38^{+0.40}_{-0.20}$ & - & -\\[5pt]
13534 & A2744 & 3.61717 & -30.42555 & $-21.80^{+0.03}_{-0.03}$ & $-2.41^{+0.08}_{-0.07}$ & $9.39^{+0.03}_{-0.03}$ & $9.32$ & Fujimoto+24\\[5pt]
25365 & A2744 & 3.59011 & -30.35974 & $-19.97^{+0.03}_{-0.03}$ & $-2.69^{+0.03}_{-0.04}$ & $10.48^{+0.10}_{-0.10}$ & $10.38$ & Fujimoto+24\\[5pt]
33044 & A2744 & 3.51193 & -30.37186 & $-20.49^{+0.03}_{-0.03}$ & $-2.13^{+0.09}_{-0.07}$ & $10.39^{+0.08}_{-0.10}$ & $9.88$ & Napolitano+25\\[5pt]
33292 & A2744 & 3.62951 & -30.37214 & $-19.16^{+0.07}_{-0.07}$ & $-2.47^{+0.08}_{-0.10}$ & $9.25^{+0.13}_{-0.13}$ & - & -\\[5pt]
37210 & A2744 & 3.56707 & -30.37787 & $-18.61^{+0.05}_{-0.04}$ & $-2.29^{+0.09}_{-0.08}$ & $10.27^{+0.18}_{-0.18}$ & $10.06$ & Fujimoto+24\\[5pt]
49489 & A2744 & 3.59250 & -30.40146 & $-18.12^{+0.03}_{-0.03}$ & $-2.52^{+0.05}_{-0.05}$ & $9.85^{+0.11}_{-0.09}$ & $9.80$ & Fujimoto+24\\[5pt]
55963 & A2744 & 3.45136 & -30.32072 & $-19.98^{+0.05}_{-0.04}$ & $-2.66^{+0.04}_{-0.04}$ & $10.40^{+0.17}_{-0.18}$ & $10.40$ & Napolitano+25\\[5pt]
56436 & A2744 & 3.45142 & -30.32181 & $-20.68^{+0.04}_{-0.03}$ & $-2.64^{+0.03}_{-0.03}$ & $10.57^{+0.13}_{-0.14}$ & $10.25$ & Napolitano+25\\[5pt]
67453 & A2744 & 3.47874 & -30.34553 & $-19.89^{+0.04}_{-0.04}$ & $-2.54^{+0.08}_{-0.08}$ & $10.67^{+0.13}_{-0.15}$ & $10.14$ & Napolitano+25\\[5pt]
11360 & COSMOS & 150.11656 & 2.19708 & $-20.96^{+0.05}_{-0.05}$ & $-2.59^{+0.07}_{-0.05}$ & $9.86^{+0.15}_{-0.14}$ & $9.81$ & Donnan+25\\[5pt]
44352 & COSMOS & 150.16258 & 2.30920 & $-20.53^{+0.07}_{-0.07}$ & $-2.48^{+0.08}_{-0.07}$ & $9.60^{+0.27}_{-0.27}$ & - & -\\[5pt]
50134 & COSMOS & 150.07230 & 2.23787 & $-19.69^{+0.06}_{-0.06}$ & $-2.33^{+0.08}_{-0.08}$ & $9.70^{+0.26}_{-0.19}$ & - & -\\[5pt]
67414 & COSMOS & 150.07313 & 2.25605 & $-20.10^{+0.04}_{-0.04}$ & $-2.27^{+0.08}_{-0.09}$ & $10.32^{+0.23}_{-0.19}$ & - & -\\[5pt]
82389 & COSMOS & 150.12637 & 2.38378 & $-20.99^{+0.05}_{-78.01}$ & $-2.51^{+0.06}_{-96.49}$ & $8.90^{+0.03}_{-0.03}$ & $8.87$ & CAPERS(6368)\\[5pt]
91271 & COSMOS & 150.16234 & 2.35739 & $-20.62^{+0.09}_{-0.10}$ & $-2.17^{+0.10}_{-0.10}$ & $9.42^{+0.41}_{-0.27}$ & - & -\\[5pt]
18031 & EGS & 214.81185 & 52.73711 & $-20.39^{+0.07}_{-0.07}$ & $-2.18^{+0.12}_{-0.11}$ & $9.94^{+0.30}_{-0.30}$ & $9.57$ & Arrabal Haro+23\\[5pt]
38537 & EGS & 214.77282 & 52.74086 & $-20.56^{+0.07}_{-0.07}$ & $-2.48^{+0.10}_{-0.07}$ & $9.33^{+0.07}_{-0.12}$ & - & -\\[5pt]
62111 & EGS & 214.99440 & 52.98938 & $-20.16^{+0.07}_{-0.06}$ & $-2.56^{+0.06}_{-0.06}$ & $9.19^{+0.11}_{-0.11}$ & $8.80$ & Fujimoto+23\\[5pt]
78312 & EGS & 214.97736 & 52.92650 & $-20.57^{+0.07}_{-0.06}$ & $-2.35^{+0.07}_{-0.05}$ & $9.23^{+0.14}_{-0.10}$ & $9.04$ & RUBIES(4233)\\[5pt]
85166 & EGS & 214.73253 & 52.75809 & $-20.41^{+0.08}_{-0.08}$ & $-2.60^{+0.06}_{-0.05}$ & $10.14^{+0.25}_{-0.28}$ & $9.85$ & Arrabal Haro+23\\[5pt]
20969 & GOODS-N & 189.33740 & 62.18073 & $-19.89^{+0.04}_{-0.04}$ & $-2.45^{+0.05}_{-0.07}$ & $9.22^{+0.08}_{-0.07}$ & - & -\\[5pt]
31580 & GOODS-N & 189.21769 & 62.19949 & $-20.15^{+0.04}_{-0.04}$ & $-2.55^{+0.06}_{-0.09}$ & $9.80^{+0.15}_{-0.12}$ & $9.75$ & JADES(1181)\\[5pt]
32549 & GOODS-N & 189.26201 & 62.20109 & $-19.91^{+0.07}_{-0.06}$ & $-2.53^{+0.06}_{-0.06}$ & $9.58^{+0.17}_{-0.14}$ & - & -\\[5pt]
38008 & GOODS-N & 189.23980 & 62.21083 & $-19.94^{+0.04}_{-0.04}$ & $-2.67^{+0.04}_{-0.05}$ & $10.18^{+0.14}_{-0.15}$ & $9.64$ & JADES(1181)\\[5pt]
38123 & GOODS-N & 189.23911 & 62.21093 & $-20.02^{+0.06}_{-0.05}$ & $-2.57^{+0.05}_{-0.04}$ & $10.05^{+0.12}_{-0.12}$ & $9.78$ & JADES(1181)\\[5pt]
38138 & GOODS-N & 189.23923 & 62.21097 & $-19.71^{+0.08}_{-0.06}$ & $-2.56^{+0.07}_{-0.05}$ & $9.57^{+0.26}_{-0.19}$ & - & -\\[5pt]
44181 & GOODS-N & 189.15825 & 62.22136 & $-20.25^{+0.06}_{-0.05}$ & $-2.50^{+0.06}_{-0.05}$ & $9.37^{+0.25}_{-0.09}$ & $9.07$ & JADES(1181)\\[5pt]
49123 & GOODS-N & 189.28903 & 62.22904 & $-19.63^{+0.06}_{-0.06}$ & $-1.90^{+0.12}_{-0.13}$ & $9.84^{+0.22}_{-0.17}$ & - & -\\[5pt]
49306 & GOODS-N & 189.28879 & 62.22930 & $-20.31^{+0.05}_{-0.05}$ & $-2.50^{+0.07}_{-0.07}$ & $10.07^{+0.20}_{-0.18}$ & - & -\\[5pt]
56195 & GOODS-N & 189.34475 & 62.23955 & $-20.97^{+0.04}_{-0.04}$ & $-2.27^{+0.06}_{-0.06}$ & $9.57^{+0.18}_{-0.16}$ & - & -\\[5pt]
58504 & GOODS-N & 189.21772 & 62.31178 & $-20.46^{+0.05}_{-0.05}$ & $-2.26^{+0.04}_{-0.04}$ & $9.53^{+0.16}_{-0.17}$ & - & -\\[5pt]
68946 & GOODS-N & 189.13833 & 62.28986 & $-20.42^{+0.05}_{-0.04}$ & $-2.64^{+0.05}_{-0.04}$ & $9.27^{+0.10}_{-0.10}$ & $9.31$ & JADES(1181)\\[5pt]
71038 & GOODS-N & 189.33650 & 62.28727 & $-20.75^{+0.03}_{-0.03}$ & $-2.61^{+0.04}_{-0.03}$ & $10.82^{+0.10}_{-0.08}$ & - & -\\[5pt]
99271 & GOODS-N & 189.10606 & 62.24205 & $-22.08^{+0.02}_{-0.02}$ & $-2.49^{+0.05}_{-0.05}$ & $10.76^{+0.06}_{-0.06}$ & $10.61$ & Bunker+23\\[5pt]
20537 & GOODS-S & 53.13696 & -27.92232 & $-19.13^{+0.08}_{-0.07}$ & $-2.69^{+0.06}_{-0.04}$ & $8.98^{+0.11}_{-0.12}$ & - & -\\[5pt]
20594 & GOODS-S & 53.17551 & -27.78064 & $-19.92^{+0.03}_{-0.03}$ & $-2.30^{+0.08}_{-0.08}$ & $9.59^{+0.07}_{-0.07}$ & $9.71$ & JADES(1180)\\[5pt]
20793 & GOODS-S & 53.21019 & -27.78115 & $-19.24^{+0.10}_{-0.12}$ & $-2.59^{+0.08}_{-0.05}$ & $9.82^{+0.34}_{-0.54}$ & - & -\\[5pt]
21022 & GOODS-S & 53.14297 & -27.92118 & $-19.42^{+0.07}_{-0.07}$ & $-2.38^{+0.07}_{-0.06}$ & $9.71^{+0.22}_{-0.20}$ & - & -\\[5pt]
21250 & GOODS-S & 53.02681 & -27.89442 & $-19.72^{+0.05}_{-0.08}$ & $-2.50^{+0.05}_{-0.06}$ & $10.33^{+0.22}_{-0.33}$ & - & -\\[5pt]
23803 & GOODS-S & 53.05582 & -27.87718 & $-19.79^{+0.03}_{-0.04}$ & $-2.37^{+0.08}_{-0.07}$ & $10.62^{+0.20}_{-0.19}$ & - & -\\[5pt]
30383 & GOODS-S & 53.05999 & -27.88514 & $-18.53^{+0.06}_{-0.07}$ & $-2.60^{+0.08}_{-0.07}$ & $10.17^{+0.20}_{-0.26}$ & - & -\\[5pt]
32068 & GOODS-S & 53.05177 & -27.88727 & $-19.13^{+0.04}_{-0.04}$ & $-2.69^{+0.07}_{-0.05}$ & $10.31^{+0.12}_{-0.17}$ & - & -\\[5pt]
37256 & GOODS-S & 53.04885 & -27.89435 & $-18.20^{+0.13}_{-0.10}$ & $-2.59^{+0.10}_{-0.08}$ & $9.62^{+0.27}_{-0.37}$ & - & -\\[5pt]
40653 & GOODS-S & 53.12222 & -27.89907 & $-19.99^{+0.06}_{-0.06}$ & $-2.32^{+0.06}_{-0.06}$ & $10.25^{+0.10}_{-0.12}$ & - & -\\[5pt]
40797 & GOODS-S & 53.15204 & -27.89469 & $-19.62^{+0.08}_{-0.07}$ & $-2.30^{+0.11}_{-0.11}$ & $10.41^{+0.25}_{-0.40}$ & - & -\\[5pt]
50841 & GOODS-S & 53.07477 & -27.84541 & $-18.00^{+0.12}_{-0.10}$ & $-2.68^{+0.05}_{-0.06}$ & $9.30^{+0.42}_{-0.26}$ & - & -\\[5pt]
52109 & GOODS-S & 53.11392 & -27.79585 & $-19.61^{+0.05}_{-0.04}$ & $-2.66^{+0.03}_{-0.03}$ & $9.27^{+0.08}_{-0.09}$ & $1.98$ & Bunker+24\\[5pt]
52254 & GOODS-S & 53.11059 & -27.84768 & $-18.90^{+0.12}_{-0.12}$ & $-2.59^{+0.08}_{-0.06}$ & $9.61^{+0.25}_{-0.32}$ & - & -\\[5pt]
53952 & GOODS-S & 53.11993 & -27.84640 & $-19.61^{+0.05}_{-0.05}$ & $-1.89^{+0.06}_{-0.07}$ & $10.02^{+0.03}_{-0.05}$ & $8.77$ & OASIS(5997)\\[5pt]
54919 & GOODS-S & 53.13362 & -27.84499 & $-20.84^{+0.04}_{-0.04}$ & $-1.89^{+0.10}_{-0.10}$ & $10.20^{+0.14}_{-0.11}$ & $9.06$ & JADES(1286)\\[5pt]
56288 & GOODS-S & 53.07388 & -27.79678 & $-20.13^{+0.08}_{-0.08}$ & $-2.31^{+0.06}_{-0.06}$ & $9.48^{+0.24}_{-0.21}$ & - & -\\[5pt]
58181 & GOODS-S & 53.16736 & -27.80751 & $-19.45^{+0.04}_{-0.05}$ & $-2.42^{+0.07}_{-0.09}$ & $10.03^{+0.28}_{-0.30}$ & $9.69$ & Bunker+24\\[5pt]
58229 & GOODS-S & 53.06020 & -27.85625 & $-18.57^{+0.08}_{-0.09}$ & $-2.64^{+0.05}_{-0.05}$ & $10.04^{+0.32}_{-0.35}$ & - & -\\[5pt]
58640 & GOODS-S & 53.07608 & -27.85601 & $-18.38^{+0.09}_{-0.14}$ & $-2.61^{+0.07}_{-0.06}$ & $9.73^{+0.39}_{-0.48}$ & - & -\\[5pt]
58713 & GOODS-S & 53.09748 & -27.85698 & $-18.71^{+0.08}_{-0.08}$ & $-2.67^{+0.04}_{-0.05}$ & $9.68^{+0.24}_{-0.25}$ & - & -\\[5pt]
58944 & GOODS-S & 53.09749 & -27.85677 & $-18.55^{+0.09}_{-0.09}$ & $-2.40^{+0.10}_{-0.10}$ & $9.94^{+0.30}_{-0.43}$ & - & -\\[5pt]
62915 & GOODS-S & 53.17603 & -27.81739 & $-18.72^{+0.07}_{-0.06}$ & $-1.91^{+0.08}_{-0.07}$ & $10.05^{+0.04}_{-0.06}$ & $8.68$ & JADES(1286)\\[5pt]
64456 & GOODS-S & 53.06774 & -27.86455 & $-18.63^{+0.09}_{-0.11}$ & $-2.51^{+0.06}_{-0.06}$ & $9.64^{+0.24}_{-0.38}$ & - & -\\[5pt]
70396 & GOODS-S & 53.06683 & -27.87293 & $-20.20^{+0.05}_{-0.04}$ & $-2.47^{+0.05}_{-0.06}$ & $10.71^{+0.13}_{-0.11}$ & - & -\\[5pt]
72355 & GOODS-S & 53.16593 & -27.83424 & $-19.90^{+0.05}_{-0.05}$ & $-2.52^{+0.06}_{-0.05}$ & $10.93^{+0.15}_{-0.16}$ & - & -\\[5pt]
74275 & GOODS-S & 53.06773 & -27.83787 & $-19.87^{+0.10}_{-0.07}$ & $-2.60^{+0.04}_{-0.04}$ & $9.36^{+0.39}_{-0.16}$ & $9.06$ & JADES(1286)\\[5pt]
77980 & GOODS-S & 53.07478 & -27.84489 & $-18.43^{+0.07}_{-0.10}$ & $-2.60^{+0.08}_{-0.06}$ & $10.26^{+0.31}_{-0.60}$ & - & -\\[5pt]
137 & PANO-M0214 & 34.35593 & -2.18771 & $-20.83^{+0.07}_{-0.07}$ & $-2.54^{+0.08}_{-0.07}$ & $10.84^{+0.20}_{-0.26}$ & - & -\\[5pt]
2694 & PANO-M0531 & 341.54387 & -5.48678 & $-19.88^{+0.08}_{-0.06}$ & $-2.41^{+0.13}_{-0.09}$ & $9.18^{+0.14}_{-0.17}$ & - & -\\[5pt]
5353 & PANO-M1019 & 24.73854 & -10.31359 & $-20.46^{+0.08}_{-0.09}$ & $-2.56^{+0.07}_{-0.07}$ & $9.12^{+0.23}_{-0.20}$ & - & -\\[5pt]
8761 & PANO-M2152 & 24.46821 & -21.87560 & $-20.61^{+0.05}_{-0.04}$ & $-1.92^{+0.09}_{-0.11}$ & $9.65^{+0.41}_{-0.25}$ & - & -\\[5pt]
9329 & PANO-M2152 & 24.46405 & -21.87899 & $-20.09^{+0.08}_{-0.09}$ & $-2.34^{+0.07}_{-0.06}$ & $9.67^{+0.30}_{-0.38}$ & - & -\\[5pt]
6429 & PANO-M2409 & 64.04908 & -24.16773 & $-19.70^{+0.09}_{-0.08}$ & $-2.50^{+0.08}_{-0.07}$ & $9.28^{+0.22}_{-0.20}$ & - & -\\[5pt]
3805 & PANO-P0025 & 334.23795 & 0.44949 & $-20.41^{+0.05}_{-0.05}$ & $-2.26^{+0.09}_{-0.08}$ & $10.96^{+0.18}_{-0.23}$ & - & -\\[5pt]
14395 & PANO-P0025 & 334.16874 & 0.39674 & $-19.92^{+0.07}_{-0.06}$ & $-2.31^{+0.12}_{-0.10}$ & $9.54^{+0.11}_{-0.11}$ & - & -\\[5pt]
14904 & PANO-P0025 & 334.23000 & 0.43057 & $-20.01^{+0.05}_{-0.06}$ & $-2.51^{+0.06}_{-0.07}$ & $9.13^{+0.10}_{-0.11}$ & - & -\\[5pt]
30966 & PANO-P0025 & 334.24302 & 0.38765 & $-20.05^{+0.04}_{-0.05}$ & $-2.47^{+0.05}_{-0.04}$ & $9.89^{+0.13}_{-0.16}$ & - & -\\[5pt]
31622 & PANO-P0025 & 334.25162 & 0.38665 & $-19.14^{+0.12}_{-0.09}$ & $-2.52^{+0.09}_{-0.08}$ & $9.07^{+0.30}_{-0.21}$ & - & -\\[5pt]
2872 & PANO-P0819 & 142.95974 & 8.33051 & $-21.58^{+0.07}_{-0.07}$ & $-2.40^{+0.11}_{-0.10}$ & $9.19^{+0.23}_{-0.18}$ & - & -\\[5pt]
717 & PANO-P5652 & 194.26913 & 56.87960 & $-21.33^{+0.13}_{-0.08}$ & $-2.60^{+0.10}_{-0.06}$ & $9.29^{+0.41}_{-0.21}$ & - & -\\[5pt]
4710 & UDS & 34.49539 & -5.31586 & $-21.36^{+0.07}_{-0.07}$ & $-1.89^{+0.11}_{-0.14}$ & $9.90^{+0.49}_{-0.49}$ & - & -\\[5pt]
11544 & UDS & 34.32489 & -5.34439 & $-21.30^{+0.15}_{-0.10}$ & $-2.49^{+0.05}_{-0.05}$ & $9.26^{+0.75}_{-0.29}$ & - & -\\[5pt]
18893 & UDS & 34.52095 & -5.29266 & $-20.49^{+0.07}_{-0.07}$ & $-2.06^{+0.10}_{-0.09}$ & $10.02^{+0.17}_{-0.24}$ & $9.30$ & CAPERS(6368)\\[5pt]
21510 & UDS & 34.42662 & -5.28894 & $-20.76^{+0.09}_{-0.08}$ & $-2.48^{+0.07}_{-0.08}$ & $9.11^{+0.20}_{-0.21}$ & - & -\\[5pt]
33482 & UDS & 34.43496 & -5.27112 & $-20.75^{+0.05}_{-0.05}$ & $-2.46^{+0.08}_{-0.06}$ & $9.46^{+0.09}_{-0.06}$ & $9.30$ & RUBIES(4233)\\[5pt]
52799 & UDS & 34.41358 & -5.24423 & $-20.18^{+0.07}_{-0.07}$ & $-2.39^{+0.10}_{-0.08}$ & $10.34^{+0.23}_{-0.41}$ & $10.32$ & CAPERS(6368)\\[5pt]
89565 & UDS & 34.46028 & -5.18499 & $-20.62^{+0.06}_{-0.06}$ & $-2.49^{+0.07}_{-0.09}$ & $9.28^{+0.13}_{-0.15}$ & $9.27$ & CAPERS(6368)\\[5pt]
118503 & UDS & 34.30583 & -5.15434 & $-20.32^{+0.05}_{-0.04}$ & $-2.48^{+0.08}_{-0.07}$ & $9.09^{+0.12}_{-0.11}$ & $9.25$ & RUBIES(4233)\\[5pt]
\hline
\end{longtable*}

\LTcapwidth=\textwidth
\begin{longtable*}{ccccccccc}

\caption{Same as Table \ref{tab:z10_candidates}, but for the $z\sim13$ F150W dropout candidates.}

\label{tab:z13_candidates}

\\& & & & & & & &\\[\dimexpr-\normalbaselineskip+0pt]
\hline
& & & & & & & &\\[\dimexpr-\normalbaselineskip+3pt]
ID & Field & R.A. & Decl. & M$_{\rm UV}$ & UV-slope $\beta$ & \shortstack{z$_{\rm phot,\,bp}$} & z$_{\rm spec}$ (DJA) & z$_{\rm spec}$ Reference\\
& & & & & & & &\\[\dimexpr-\normalbaselineskip+3pt]
\hline
\hline
57585 & A2744 & 3.49898 & -30.32476 & $-20.82^{+0.04}_{-0.04}$ & $-2.54^{+0.08}_{-0.07}$ & $12.47^{+0.08}_{-0.07}$ & $12.34$ & Castellano+24\\[5pt]
13670 & COSMOS & 150.06116 & 2.16076 & $-20.01^{+0.07}_{-0.10}$ & $-2.35^{+0.11}_{-0.10}$ & $13.29^{+0.44}_{-0.69}$ & - & -\\[5pt]
58174 & COSMOS & 150.09917 & 2.24795 & $-20.05^{+0.11}_{-0.09}$ & $-2.13^{+0.13}_{-0.10}$ & $12.29^{+0.63}_{-0.35}$ & - & -\\[5pt]
1682 & GOODS-S & 53.01081 & -27.74203 & $-20.83^{+0.11}_{-0.08}$ & $-2.04^{+0.11}_{-0.12}$ & $11.98^{+0.37}_{-0.25}$ & - & -\\[5pt]
37941 & GOODS-S & 53.04019 & -27.89644 & $-19.70^{+0.08}_{-0.07}$ & $-1.92^{+0.07}_{-0.08}$ & $12.74^{+0.57}_{-0.40}$ & - & -\\[5pt]
66144 & GOODS-S & 53.16634 & -27.82156 & $-18.92^{+0.12}_{-0.09}$ & $-2.50^{+0.10}_{-0.10}$ & $12.10^{+0.47}_{-0.28}$ & $12.51$ & Curtis-Lake+23\\[5pt]
69545 & GOODS-S & 53.04683 & -27.87190 & $-18.85^{+0.16}_{-0.11}$ & $-2.25^{+0.12}_{-0.11}$ & $12.41^{+1.22}_{-0.45}$ & - & -\\[5pt]
84448 & GOODS-S & 53.08294 & -27.85563 & $-21.29^{+0.02}_{-0.02}$ & $-2.22^{+0.05}_{-0.06}$ & $14.19^{+0.07}_{-0.06}$ & $14.18$ & Carniani+24\\[5pt]
35840 & PANO-P0025 & 334.25036 & 0.37920 & $-21.18^{+0.04}_{-0.04}$ & $-2.54^{+0.10}_{-0.09}$ & $14.55^{+0.16}_{-0.18}$ & $13.53$ & Donnan+26\\[5pt]
3640 & PANO-P5549 & 205.94732 & 55.80276 & $-21.72^{+0.06}_{-0.06}$ & $-2.43^{+0.16}_{-0.10}$ & $14.19^{+0.24}_{-0.36}$ & - & -\\[5pt]
\hline
\end{longtable*}

\textbf{}
\clearpage
\end{document}